\documentclass[fleqn,usenatbib]{mnras}

\usepackage{newtxtext,newtxmath}

\usepackage[T1]{fontenc}
\usepackage{ae,aecompl}

\usepackage{graphicx}	
\usepackage{amsmath}	
\usepackage[dvipsnames]{xcolor}
\usepackage{physics}
\usepackage[normalem]{ulem}
\newcommand{\dt}{$\Delta\tau \:$}
\usepackage{lineno}
\DeclareUnicodeCharacter{2212}{-}
\usepackage{cleveref}
\creflabelformat{equation}{#2#1#3}

\let\oldequation\equation
\let\oldendequation\endequation

\renewenvironment{equation}
  {\linenomathNonumbers\oldequation}
  {\oldendequation\endlinenomath}

\newcommand{\black}{\color{black}}

\usepackage{eso-pic}

\AddToShipoutPictureBG*{%
  \AtPageUpperLeft{%
    \hspace{0.75\paperwidth}%
    \raisebox{-5\baselineskip}{%
      \makebox[0pt][l]{\textnormal{DES-2020-0539}}
}}}%

\AddToShipoutPictureBG*{%
  \AtPageUpperLeft{%
    \hspace{0.75\paperwidth}%
    \raisebox{-6\baselineskip}{%
      \makebox[0pt][l]{\textnormal{FERMILAB-PUB-21-003-AE}}
}}}%

\title[OzDES Lag Recovery Techniques]{OzDES Reverberation Mapping Program: Lag recovery reliability for 6-year C~\textsc{iv} analysis}

\author[A. Penton et al.]{
\parbox{\textwidth}{
\Large
A.~Penton$^{1},$\thanks{E-mail: a.penton@uq.edu.au (UQ)}
U.~Malik,$^{2}$
T.~M.~Davis,$^{1}$
P.~Martini,$^{3,4,5}$
Z.~Yu,$^{4}$
R.~Sharp,$^{2}$
C.~Lidman,$^{6,2}$
B.~E.~Tucker,$^{2}$
J.~K.~Hoormann,$^{1}$
M.~Aguena,$^{7,8}$
S.~Allam,$^{9}$
J.~Annis,$^{9}$
J.~Asorey,$^{10}$
D.~Bacon,$^{11}$
E.~Bertin,$^{12,13}$
S.~Bhargava,$^{14}$
D.~Brooks,$^{15}$
J.~Calcino,$^{1}$
A.~Carnero~Rosell,$^{16}$
D.~Carollo,$^{17}$
M.~Carrasco~Kind,$^{18,19}$
J.~Carretero,$^{20}$
M.~Costanzi,$^{21,22}$
L.~N.~da Costa,$^{8,23}$
M.~E.~S.~Pereira,$^{24}$
J.~De~Vicente,$^{10}$
H.~T.~Diehl,$^{9}$
T.~F.~Eifler,$^{25,26}$
S.~Everett,$^{27}$
I.~Ferrero,$^{28}$
P.~Fosalba,$^{29,30}$
J.~Frieman,$^{9,31}$
J.~Garc\'ia-Bellido,$^{32}$
E.~Gaztanaga,$^{29,30}$
D.~W.~Gerdes,$^{33,24}$
D.~Gruen,$^{34,35,36}$
R.~A.~Gruendl,$^{18,19}$
J.~Gschwend,$^{8,23}$
G.~Gutierrez,$^{9}$
S.~R.~Hinton,$^{1}$
D.~L.~Hollowood,$^{27}$
K.~Honscheid,$^{3,37}$
D.~J.~James,$^{38}$
A.~G.~Kim,$^{39}$
K.~Kuehn,$^{40,41}$
N.~Kuropatkin,$^{9}$
M.~A.~G.~Maia,$^{8,23}$
J.~L.~Marshall,$^{42}$
F.~Menanteau,$^{18,19}$
R.~Miquel,$^{43,20}$
R.~Morgan,$^{44}$
A.~M\"oller,$^{45}$
A.~Palmese,$^{9,31}$
F.~Paz-Chinch\'{o}n,$^{18,46}$
A.~A.~Plazas,$^{47}$
A.~K.~Romer,$^{14}$
E.~Sanchez,$^{10}$
V.~Scarpine,$^{9}$
D.~Scolnic,$^{48}$
S.~Serrano,$^{29,30}$
M.~Smith,$^{49}$
E.~Suchyta,$^{50}$
M.~E.~C.~Swanson,$^{18}$
G.~Tarle,$^{24}$
C.~To,$^{34,35,36}$
S.~A.~Uddin,$^{51}$
T.~N.~Varga,$^{52,53}$
W.~Wester,$^{9}$
and R.D.~Wilkinson$^{14}$
and (DES Collaboration)
\\ 
{\it \footnotesize Affiliations are listed at the end of the paper}
}
}

\date{Accepted 2021 October 14. Received 2021 October 14; in original form 2021 January 18}

\pubyear{2021}

\begin{document}
\label{firstpage}
\pagerange{\pageref{firstpage}--\pageref{lastpage}}
\maketitle

\begin{abstract}
We present the statistical methods that have been developed to analyse the OzDES reverberation mapping sample. To perform this statistical analysis we have created a suite of customisable simulations that mimic the characteristics of each source in the OzDES sample. These characteristics include: the variability in the photometric and spectroscopic lightcurves, the measurement uncertainties, and the observational cadence. By simulating the sources in the OzDES sample that contain the \ion{C}{iv} emission line, we developed a set of criteria that rank the reliability of a recovered time lag depending on the agreement between different recovery methods, the magnitude of the uncertainties, and the rate at which false positives were found in the simulations. These criteria were applied to simulated light curves and these results used to estimate the quality of the resulting Radius-Luminosity relation.We grade the results using three quality levels (gold, silver and bronze). The input slope of the \textit{R-L} relation was recovered within $1\sigma$ for each of the three quality samples, with the gold standard having the lowest dispersion with a recovered a \textit{R-L} relation slope of $0.454\pm 0.016$ with an input slope of 0.47.  Future work will apply these methods to the entire OzDES sample of 771 AGN.
\end{abstract}



\section{Introduction}\label{intro}

The innermost regions of active galactic nuclei (AGN) are powered by supermassive black holes, whose role in galaxy formation and evolution is complex and poorly understood. For AGN within the local Universe, high spatial resolution instruments are capable of probing the sphere of influence of the central black hole and directly measuring the mass \citep[e.g.,][]{Gebhardt2000,Greene2010,Gebhardt2011,Kuo2011,EHT2019}. However, we require alternate methods to study AGN at greater distances in order to explore the evolution of supermassive black holes. For this purpose, Reverberation Mapping (RM) can be used to directly measure distances within these compact regions and infer the masses of the central supermassive black holes (SMBH). 

The technique of Reverberation Mapping \citep{Blandford1982, Peterson1993} uses time-domain observations to provide a window to AGN physics on spatial scales below the angular resolution of contemporary observatories. The prompt and variable rest-frame UV emission from the accretion disk ionises the more extended broad-line region (BLR). Variations in the UV continuum radiation from the disk produce a variation in the observed emission-line signal over an extended time scale, on the order of months to years. The observed reverberation of the BLR in response to the UV continuum is due to the light crossing time from the central source to the BLR and the geometry of the BLR. Therefore, by measuring this time delay, $\tau$, we can measure of the radius of the BLR ($R_{\textrm{BLR}} = c \tau$). The velocity dispersion of the BLR ($\Delta V$) can be estimated from the width of the broadened emission lines. The mass of the central black hole ($M_{\textrm{BH}}$) can then be measured using the Virial theorem:
\begin{linenomath}
    \begin{equation}\label{virial}
        M_{\textrm{BH}} = f\frac{R_{\textrm{BLR}} \Delta V^2}{G},
    \end{equation}
\end{linenomath}
where $f$ is the virial coefficient; a dimensionless scale factor that accounts for the geometry, orientation, and kinematics of the BLR.

Extensive time-domain monitoring of both the continuum emission and emission line flux is required to conduct reverberation mapping of the BLR. Due to limits of technology at the time, early campaigns targeted few bright, highly-variable sources, which corresponded to relatively low-luminosity AGN in the local Universe. Subsequent generations of these surveys over many years gradually produced a sample of reliable lag measurements for 63 AGN \citep[e.g.,][]{Kaspi2000,Onken2002,Peterson2004,Bentz2009b,Denney2010,Barth2011,Grier2012,Bentz2015}.As most of these sources were at low redshifts ($z<0.3$), most results were obtained using the H$\beta$ emission line. The observations confirmed the predicted relationship between the AGN luminosity and the radius of the BLR \citep{Kaspi2000,Bentz2006,Bentz2009a}. 

This Radius-Luminosity ($\textit{R-L}$) relationship is a powerful tool to estimate SMBH masses from a single-epoch spectroscopic measurement. This has allowed single-epoch Virial BH mass estimates to be made for tens of thousands of objects \citep{shen2011}, in order to study SMBH evolution. However, for sources at higher redshifts (and hence greater evolutionary lookback times), H$\beta$ is redshifted into the near-infrared spectrum and becomes increasing challenging to observe. For these more distant sources, both single-epoch and RM observations rely on emission from Mg~\textsc{ii} and C~\textsc{iv}, for which a detailed $\textit{R-L}$ relation calibration is not yet available. This inhibits the direct construction of single-epoch virial BH estimates for these important sources. Single-epoch SMBH mass estimators  based on C~\textsc{iv} are calibrated based on UV spectra of local sources \citep{vestergaard2006}.

These AGN have longer lags, due to both the increased intrinsic luminosity of the observed sources, and the impact of time dilation, thus requiring long-baseline monitoring. With the C~\textsc{iv} line, significant RM measurements have been made for an additional 65 AGN to date \citep{Peterson2004,Peterson2005,Metzroth2006,Kaspi2007,Trevese2014,Lira2018,Hoormann2019,Grier2019,2019SDSS}.

\begin{figure*}
    \centering
    \includegraphics[trim={0cm 1cm 0cm 0cm},clip,scale=0.7]{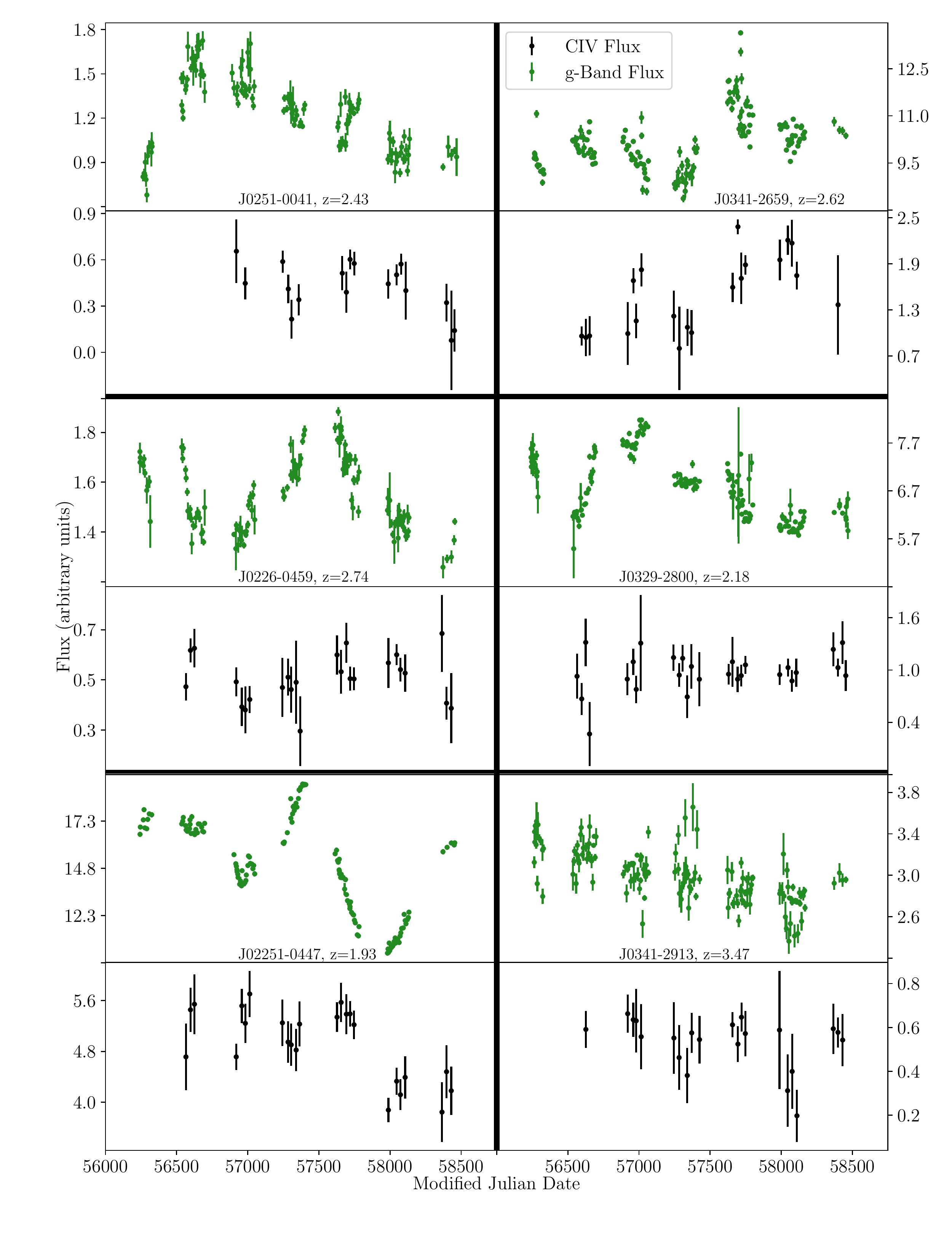}
    \caption{Representative lightcurves from the OzDES RM sample. \ion{C}{iv} emission line lightcurves shown in black, with g-band photometric lightcurves shown in green.  The photometric lightcurves contain 7 seasons of data with approximately weekly cadence. These are accompanied by spectroscopic lightcurves containing 5-6 seasons of approximately monthly cadence. }
    \label{fig:lightcurves}
\end{figure*}

Recent `industrial-scale' Reverberation Mapping campaigns have probed new regions of the AGN luminosity-redshift parameter space, with a particular focus on  high-redshift sources. The Australian Dark Energy Survey \citep[OzDES; see][]{Yuan2015, Childress2017} began one of the first multi-object RM campaigns, monitoring 771 AGN over a 6-year baseline with the Anglo-Australian Telescope. This was complimented by photometric monitoring of the same sources in the Dark Energy Survey \citep[DES; see][]{DES2016MNRAS.460.1270D} deep fields for the same time period (see \Cref{fig:lightcurves}). With the ability to conduct multi-object spectroscopy, OzDES was able to targets hundreds of AGN over a broad range of redshifts (0.1 $<z<$ 4.5) and luminosity (apparent $r$-band AB magnitudes from 17 $<r<$ 22.5). About one-third of these AGN are at redshifts greater than 1.7, where the C~\textsc{iv} line is visible. \cite{Hoormann2019} published our first RM results with the C~\textsc{iv} line, for two sources at redshifts of 1.905 and 2.593, which are among the highest redshift and highest mass black holes measured to date with RM.

Due to our goal of measuring these high-redshift long-duration AGN time-lags, the observational window of our survey differed from traditional RM programs that employ single-object spectrographs. A multi-year baseline was required to ensure the longer lags could be measured. As the spectroscopic counterpart of the Dark Energy Survey, we monitored the supernova fields \citep{DESobs2019}, which were visible for $\sim$6 months of the year. We used a lower cadence for the spectroscopic monitoring than traditional surveys. Monthly monitoring of an AGN at $z\sim3$ is approximately equivalent to weekly monitoring of an AGN at $z\sim0.1$ because of the factor of $\sim4$ in time dilation. A similar industrial-scale survey was conducted by the Sloan Digital Sky Survey Reverberation Mapping Project (SDSS-RM) \citep{Shen2015}. Simulations for the OzDES RM and SDSS-RM programs \citep{King2015,Shen2015} and recent RM results from these programs \citep{Hoormann2019, Grier2019,2017Hbeta,2020MgII} show how the observational window presents challenges for recovery of these high-$z$ AGN lags, such as aliasing due to seasonal gaps. In addition, lag recovery depends on the signal-to-noise of the flux measurements and observed intrinsic continuum variability of the AGN. We were motivated to develop more sophisticated statistical techniques by the complications of seasonal gaps, changes of cadence, and variations in S/N of both the continuum and emission line data.

The most widely used lag recovery methods in the literature are the Interpolated Cross-Correlation Function (ICCF),\citep{Gaskell1987} and \textsc{javelin} \citep{Zu2011}. These techniques have proven to recover reliable and consistent lags for traditional RM surveys, however this has not conclusively been shown for large-scale RM programs targeting higher-$z$ AGN. The restricted signal-to-noise and more limited sampling of these programs dictate a rigorous analysis of the biases and false positive rates, to devise robust lag recovery and confidence criteria. Two comprehensive studies comparing lag recovery methods have been performed recently. Simulations conducted by \citet{Li2019}, specifically for SDSS-RM, found \textsc{javelin} performed better overall than ICCF, but were performed with preset detection criteria on populations of sources, rather than the individually customised simulations that will be used here to inform our significance criteria. These results were corroborated by \citet{Yu2020}, who found \textsc{javelin} produced more correct lag uncertainties, however their results were based off simulated light curves of a few local sources that had been monitored at very high cadence.

In this work, we conduct simulations using mock light curves representative of our data quality, created on a source-by-source basis, to compare the performance of these lag recovery techniques. This is used to determine the recovery and significance criteria that will be used for following OzDES RM analyses with each emission line. For each source on the OzDES RM \ion{C}{iv} catalogue, a set of bespoke simulations will be run using the observable parameters for that source while letting not observable characteristic such as time lag and black hole mass vary. This will be discussed in full in \Cref{Simulations}. The lag recovery methods and structure of our simulations are discussed in \Cref{lag_rec}. In \Cref{selection} we outline our quality criteria and apply these cuts to analyse the resulting \textit{R-L} relations in \Cref{rldata}. We summarise our results and outlook to the future in \Cref{summary}.

\section{Simulations}\label{Simulations}

Simulations have become an important part of assessing the accuracy of reverberation mapping lag recoveries. Before wide-scale simulations were computationally easy, reverberation mapping studies used other means to gauge the statistical reliability of their lags -- such as using the number of negative lags ($\tau<0$) as a measure of the expected false positive recovery rate \citep{Grier2019}.\footnote{ Since negative lags are unphysical \citep{Grier2019} assumes they are all spurious, and calculates the false positive rate on the assumption that there will be as many random false positives with $\tau>0$ as $\tau<0$.}  Since then simulations have been introduced as a means to improve that estimation of uncertainties.  The largest simulation suite to date was run by \citet{Li2019}, who simulated a large variety of mock sources that spanned the observational features (redshift, luminosity, etc. ) of their data.  In this work we go one step further, and make bespoke simulations for {\em each individual source} in our sample.

We use the same AGN variability model as \citet{Li2019}. This is based on \citet{Kelly2009} who showed that a damped random walk (DRW) can be used to model the stochastic variability of AGN light curves. A DRW is a random walk with an additional term that pushes deviations back to the mean value. It is characterised by two parameters, the damping timescale and the amplitude, which are unique to the source. \citet{Kozlowski2010} and \citet{MacLeod2010} extended this work and compared this model to more observed AGN light curves, applying the DRW model to directly constrain the variability parameters. \citet{MacLeod2010} determined the correlations between the variability parameters and physical AGN properties, including luminosity and black hole mass, using photometric light curves for $\sim$8,000 spectroscopically confirmed quasars in the Stripe 82 field, which were monitored over a 10 year baseline by the Sloan Digital Sky Survey (SDSS).

We simulate light curves following the method described by \citet{King2015}, which is the same DRW model used by \citet{Kelly2009}, \citet{Kozlowski2010}, and \citet{MacLeod2010}, applied specifically for each of the objects in the OzDES RM program. The continuum and emission-line light curves are created as described in the \Cref{lightcurves}. Following this, we describe the customisation for each source, and the simulation set-up used for our analysis.

\subsection{Continuum and emission-line light curves} \label{lightcurves}

The following parameters are required to model the continuum and emission-line light curves for an AGN:
\begin{itemize}
\setlength{\itemsep}{-0.1\baselineskip}
\item mean of the lightcurve, $\mu$;
\item damping timescale, $\tau_D$, in days;
\item long-term structure function, $SF_\infty$, in mag;
\item lag, $\tau$, in days.
\end{itemize}
The damping timescale (also referred to as the relaxation time or characteristic time-scale) is the average time it takes for the random walk to return to the mean. The amplitude of the DRW can be described a function of the standard deviation of the DRW known as the structure function, $SF(\Delta t)$. The simulated light curves for the OzDES AGN sample need to be generated for a survey baseline of at least $\Delta t = 7$ years. The asymptotic value of the structure function at large $\Delta t$ is:
\begin{equation}\label{eq:SFinf}
SF(\Delta t \gg \tau_D) \equiv SF_\infty = \sqrt[]{2}\sigma
\end{equation} 
where $\sigma$ is the long-term standard deviation of the variability. 

The continuum lightcurve, in magnitudes, is defined by a damped random walk with a mean $\mu$, and variable term $\Delta C(t)$: 
\begin{equation}
C(t) = \mu +\Delta C(t)
\end{equation}
where $\mu$ is the monochromatic continuum flux density at a given wavelength, converted to an apparent magnitude. The variable term at $t = 0$ is $\Delta C(t_0)$ = $\sigma G(1)$, where $\sigma$ is as defined in Equation \ref{eq:SFinf}, and $G(1)$ is a random number drawn from a Gaussian distribution with a mean of 0 and standard deviation of 1. Subsequent variable terms are given by:
\begin{align}
\Delta C(t_{i+1}) = \Delta C(t_i) \exp& \left( \frac{-|t_{i+1} - t_i|}{\tau_D} \right) \\
&+ \sigma G(1) \left[1 - \exp \left( \frac{−2|t_{i+1} - t_i|}{\tau_D} \right) \right]^{\frac{1}{2}}. 
\end{align}

\citet{Blandford1982} interpret the emission-line flux variations as a response to continuum variations using: 
\begin{equation}
\Delta L (t) = \int \Psi (\tau) \Delta C (t - \tau) \textrm{d}\tau,
\end{equation}
where $\Delta L(t)$ is the emission-line light-curve flux relative to its mean value, $\Psi (\tau)$ is the transfer function, $C(t)$ is the variable component of the continuum lightcurve flux and $\tau$ is the lag. The transfer function describes the BLR emission-line flux response to a delta function variation of the continuum flux. It has the effect of smoothing the emission-line lightcurve and shifting it, relative to the continuum lightcurve, by the lag $\tau$. We convolve the continuum lightcurve with a top-hat transfer function to generate the smoothed and shifted emission-line lightcurve. As the true form of $\Psi(\tau)$ is complex and related to the geometry and kinematics of the BLR \citep{2001peterson}, we use the top-hat as an approximation. As in \citet{Zu2011}, we use a top-hat transfer function of the form:
\begin{equation}
\Psi (t) = \begin{cases}
	\displaystyle \frac{1}{w} & \tau - w/2 <  t < \tau + w/2 \\
    0 & \textrm{otherwise}
\end{cases},
\end{equation}
where $w$ is the width of the top-hat. Following \citet{King2015} we adopt $w = 0.1\tau$.  

To generate light curves for the AGN sample monitored by the OzDES RM program, the four parameters described above ($\mu, \tau, \tau_D, SF_\infty$) were used to create a bespoke simulation for each source. The parameters were found using the apparent $r$-band AB magnitudes and redshifts unique to each source, as described in \Cref{sim_param}. The light curve's magnitudes are also scaled such that their magnitudes and variations are consistent with the lightcurve of the source from which they are modelled (`parent' source).

\subsection{Cadence and uncertainties} 
\begin{figure}
    \centering
    \includegraphics[scale=0.55]{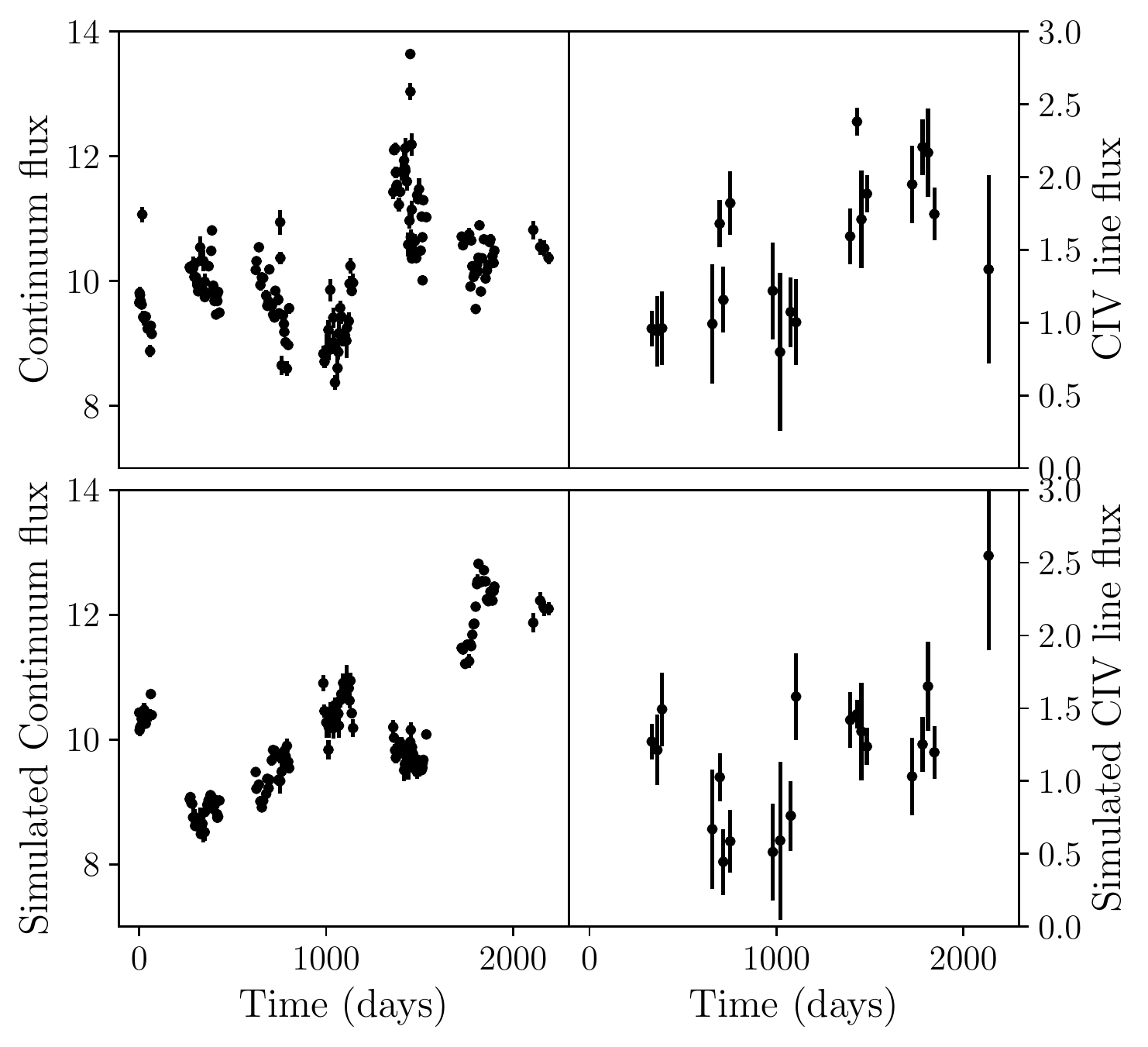}
    \caption{Comparison of observed continuum and emission line lightcurves (top) to simulated continuum and emission line lightcurves (bottom). Note that whilst they are inherently different, the uncertainties, cadence, mean and variability are consistent between the top and bottom panels.
     }
    \label{lightcurve_comp}
\end{figure}

Our custom simulations have the same cadence and noise properties as the data for each AGN. We construct them by producing high cadence lightcurves using the method illustrated in the previous sections. These are then subsampled to have identical cadence as their `parent' source. This ensures that any effects that are a function of the observational window are reflected in the simulations. In addition to this, the absolute errors from the parent source are used directly. This ensures that the simulated lightcurves include any observational effects caused by the survey. The final result of this process is shown in \Cref{lightcurve_comp}.

\subsection{Matching simulations to data variability}
\begin{figure}
    \centering
    \hspace*{-0.6cm}\includegraphics[scale=0.6]{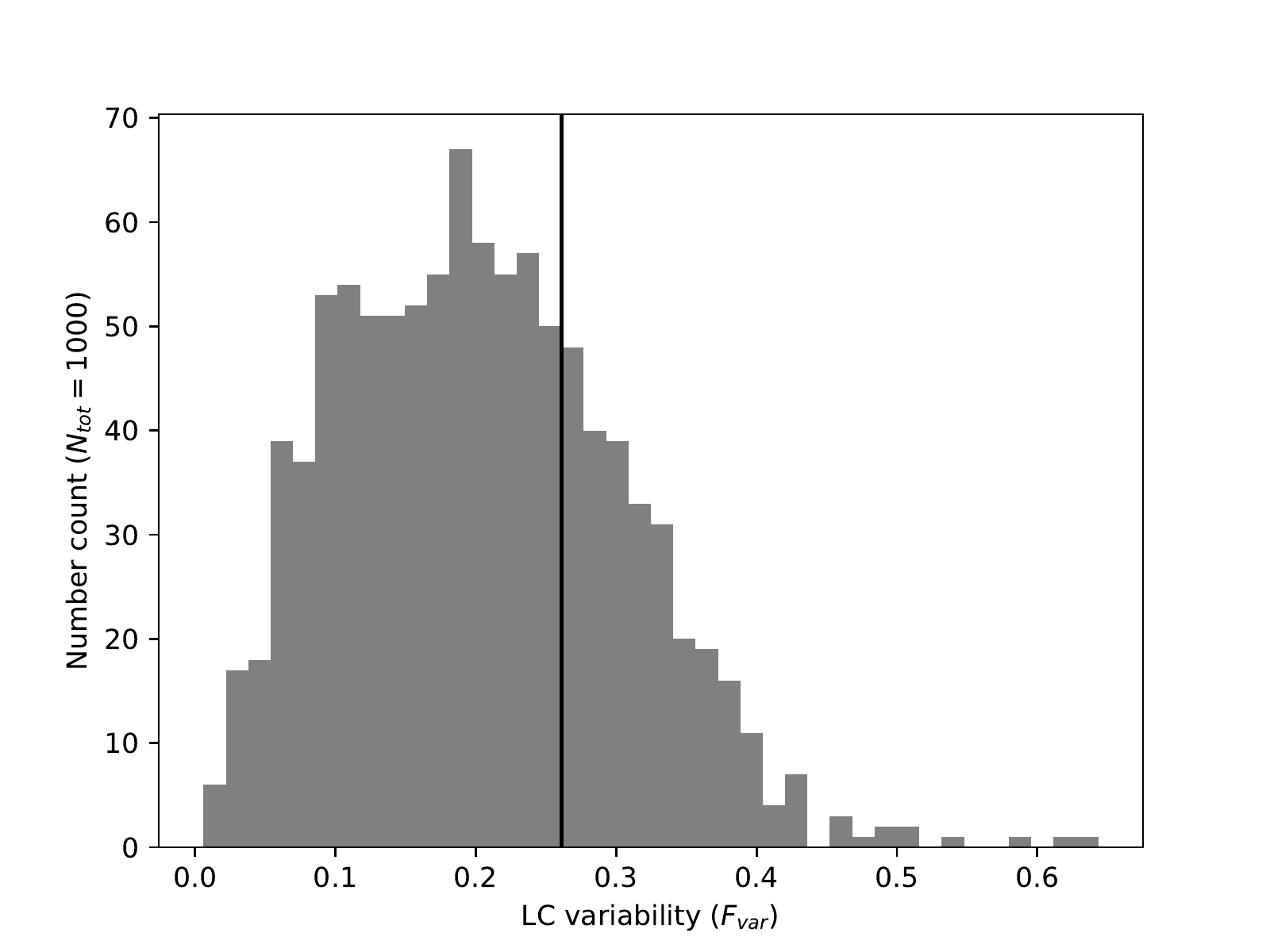}
    \caption{Variability of 1000 subsamples of the same underlying lightcurve. This shows the dispersion in the apparent variability from subsampling a simulation to match the observing cadence. The black vertical line indicates the measured variability of the parent source.}
    \label{fig:post_sel_just}
\end{figure}

Due to the sub-sampling, the difference in variability of different realisations could vary considerably by chance. This can be shown by subsampling the same underlying lightcurve with the same cadence but with different starting points. The distributions of variability are shown in \Cref{fig:post_sel_just}. Since it is likely that lightcurve variability is an important parameter in the recoverability of a time lag, it is vital that this is representative in the simulations.

To ensure that the simulated lightcurves closely match the data we perform a post-selection based on the lightcurve variability. Performing the post-selection is done by retaining the photometric-spectroscopic pair of lightcurves only if the measured variability after subsampling is within 33\% of the observed variability of the input source. In this case the variability is quantified to be the fractional variability $F_{\text{var}}$ \citep{2016ApJ...821...56F} to encapsulate the variation of the lightcurve inclusive of errors,
\begin{equation}
    \hspace*{0.5cm}F_{\text{var}} = \frac{1}{\langle f(t)\rangle}\sqrt{\frac{1}{N}\sum_{i}^{N} \left\{\left[f(t_i)-\langle f(t)\rangle\right]^2-\sigma_i^2\right\}}.
\end{equation}
Where $f(t_i)$ are flux values in the lightcurve and $\sigma_i$ are the errors on each data point.
This subsampling process also allows us to allow some freedom in our input parameters, importantly the BH mass. Assuming a specific black hole mass would likely bias the simulations as the black hole mass is not accurately known. Therefore, for each simulated lightcurve a new black hole mass is drawn from the parent distribution (\Cref{fig:BHMdist}). This allows some realisations to have a high black hole mass, and therefore a high intrinsic variability, while others have low black hole mass and a low intrinsic variability. Both can appear to have the same variability after subsampling based on observational cadence and this method allows us to not be biased to any specific black hole mass based on variability.

\subsection{Range of time-lags simulated}\label{lagfracs}
More luminous AGN tend to have longer time-lags.  One can use the $R$--$L$ relation to predict the time-lag ($\tau_{\rm expected}$) for a source based on its absolute luminosity.  However, for each of the sources that is used in this analysis, a range of different lags has been considered. This means that not only were the sources simulated with the expected time delay but with a range of seven time delays ranging from $40 \%$  of $\tau_{\rm expected}$ to $160\%$ of $\tau_{\rm expected}$, 
giving seven sets of simulations for each source, each containing 200 lightcurves, all of which pass the variability selection discussed in the previous section, giving a total of 1400 simulations per source. This was done in an attempt to not bias our analysis towards recovering lags that were exactly what were expected.

\section{Lag Recovery Methods}\label{lag_rec}
Two of the most commonly used lag recovery methods are the Interpolated Cross-Correlation Function  \citep[ICCF;][]{Gaskell1987, Peterson1998} and JAVELIN \citep{Zu2011, Zu2013}. The ICCF uses linear interpolation to provide information about the lightcurve between data points. Under the assumption of smooth variations in light-curve structure on intermediate time-scales, linear interpolation of the observational data sets maps the sparse sampled photometric and spectroscopic light curves to a common sampling frequency prior to cross-correlation. The statistical and systematic uncertainties in the cross correlation are estimated via bootstrap sampling \citep{Gaskell1987, Peterson1998}.

Employing a more sophisticated statistical model, JAVELIN utilises a Markov chain Monte Carlo (MCMC) approach based on a Damped Random Walk model (\Cref{Simulations}) for AGN variability. This is then used to constrain the time lag between lightcurves. The prior range set on the time lag search for both ICCF and JAVELIN is 0 days to $3\times \tau_{\text{exp}}$ days, where $\tau_{\text{exp}}$ is the time lag that is estimated using the known \textit{R-L} relations. While both ICCF and JAVELIN will be considered and employed in this analysis, the final results will utilise JAVELIN results.

A third contemporary lag recovery methodology, CREAM \citep{CREAM}, uses similar methods as JAVELIN to constrain the time lag, however this method is not considered in this analysis at this time. For other possible methods for recovering time-lags see \citet{2021ApJ...912...10Z}.

\subsection{Lag Posterior Analysis}\label{uncertainties}
\begin{figure}
    \centering
    \includegraphics[scale=0.59]{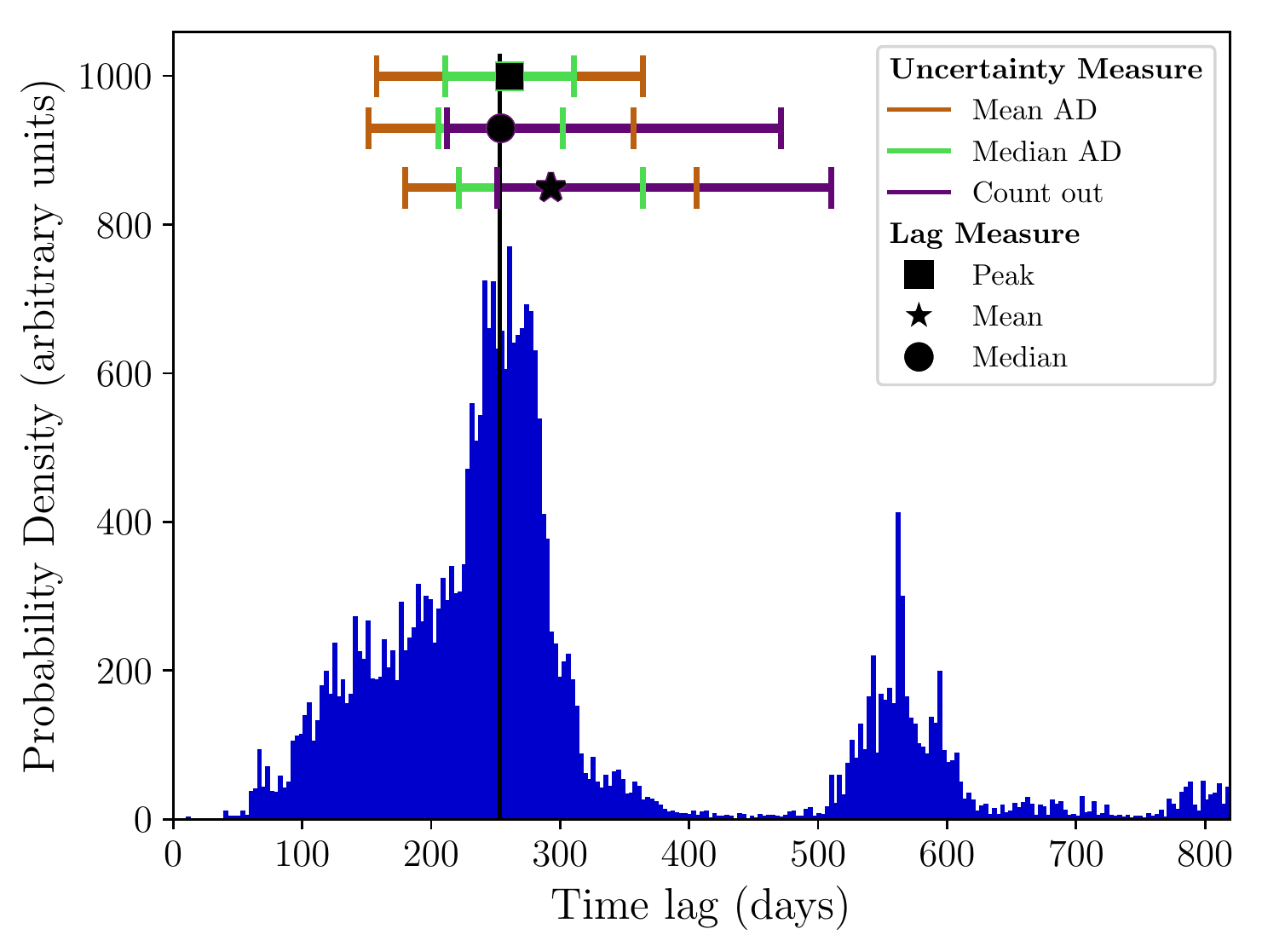}
    \caption{Example of the output PDF from JAVELIN with best fits and error bars shown using different methods of recovering lags and uncertainties. The three distinct groups represent three different ways of measuring the lag itself: the peak, the mean, and the median of the PDF. The colours in each set show the errorbars associated with using the different methods of measuring uncertainty: the mean absolute deviation, the median absolute deviation, and Count Out which employs 34\% of the probability on either side of the lag. The black vertical line marks the input lag of the simulation. Note: there is no Count Out option for the peak lag recovery method as the method often failed in noisy PDFs.}
    \label{fig:methods_comp}
\end{figure}

From both JAVELIN and ICCF lag recovery methods the output is a probability distribution function (PDF) of possible lags as seen in \Cref{fig:methods_comp}. There are multiple ways that a time lag and uncertainty can be computed from the PDF. The choice of how to compute a lag and uncertainty is important as it can vastly affect the result see \Cref{fig:methods_comp}.

Our goal is to obtain an unbiased measurement of the lag and its uncertainty. For this we must find a method that displays two important characteristics: no systematic bias and uncertainties that are the correct size. To answer this we considered three methods to determine the most representative way to determine the lag: the mean, the median, and the peak of the PDF. These methods were then applied to the PDFs  from JAVELIN and ICCF for all simulations discussed in \Cref{lagfracs}. Note that the the PDFs for both JAVELIN and ICCF have a bin width of 3 days, this was found to be the smallest bin size that made the PDFs smooth enough to accurately define the peak.\\

\begin{figure}
    \centering
    \hspace*{-0.6cm}\includegraphics[scale=0.48]{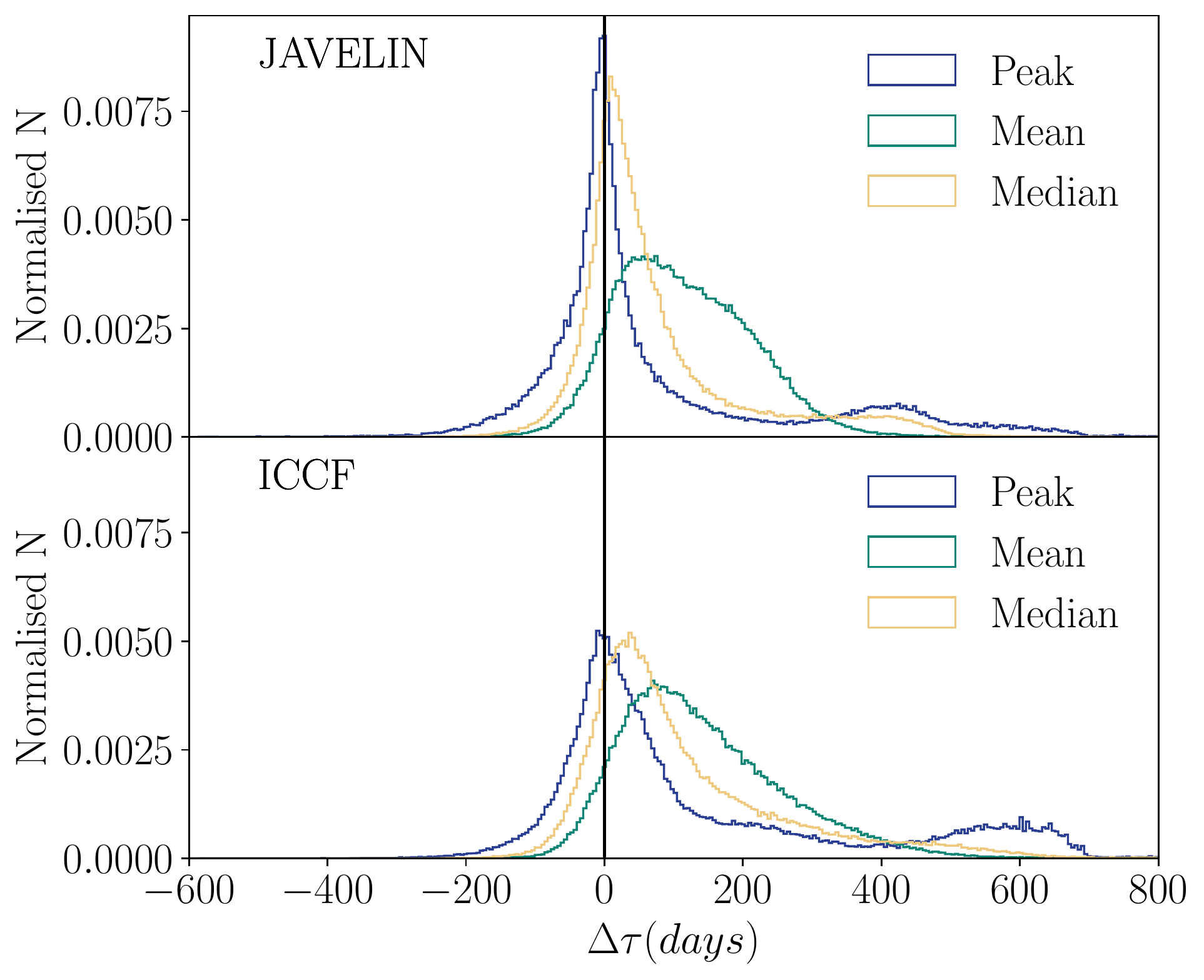}
\caption{The effectiveness of the different lag recovery methods based on whole simulation sample. The black vertical line represents a $\Delta \tau$ of 0, indicating simulations that accurately recovered their input lag. The panels showing the effectiveness of JAVELIN and ICCF display similar trends. In both, the peak measurement is the most well centered on 0, however, both show significant artifacts at positive \dt.}
    \label{fig:delta_tau_all}
\end{figure}

\begin{figure*}
    \centering
    \includegraphics[scale=0.67]{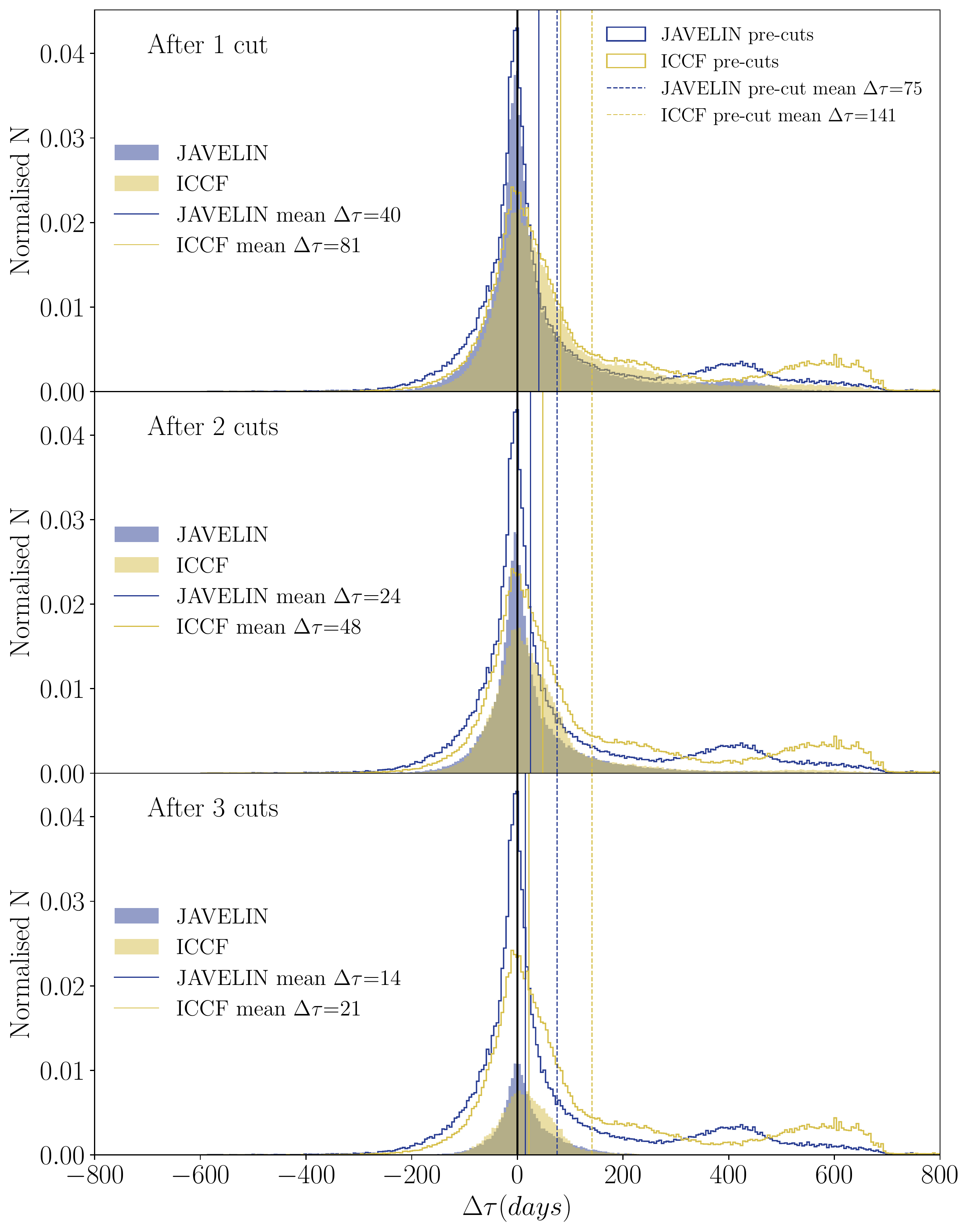}
    \caption{Distribution of $\Delta\tau$ using the peak of the JAVELIN and ICCF distributions for each of the cuts discussed \Cref{sect:quality}. \textbf{Top:} first cut, enforcing a limit of $<100$ days between the peak and median measurement for each PDF. \textbf{Center:} second cut, enforcing cut one plus a limit of $<100$ days between the JAVELIN and ICCF measurement for each PDF. \textbf{Bottom:} third cut, enforcing cut 2 plus constraining the maximum uncertainty to 80 days measurement for each lag measurement. The exact thresholds on these cuts will be discussed further in \Cref{determining_cut_size}. We can see that the artifacts present in both JAVELIN and ICCF recoveries at high $\Delta\tau \:$are reduced by each subsequent cut. This indicates that the cause of this anomaly is different for JAVELIN and ICCF as well as the peak and median measurements, and predominantly occurs in lag measurements that have larger uncertainties. This results in each cut reducing the mean offset further for both JAVELIN and ICCF.}
    \label{fig:delta_tau_all_cuts}
\end{figure*}

The difference between the recovered lag and the simulation's input lag ($\Delta\tau=\tau_{\text{sim}}-\tau_{\text{true}}$) should ideally be zero. \Cref{fig:delta_tau_all} compares the $\Delta\tau$ distributions for the three different methods for computing the time lag, for the simulations of our entire sample. From this we conclude that the mean appears to be the poorest estimator of time lag, giving a positively skewed distribution $\Delta \tau$. This is likely due to poorly constrained lags having means that are central to the prior range. Since in this case the prior range is zero to three times the input lag, the mean estimate will often skew upwards since the center of the prior range is greater than then input lag. Once the poorly constrained lags are removed with cuts outlined in \Cref{sect:quality}, this problem is greatly reduced. 

\subsection{Quality cuts}\label{sect:quality}
Using the peak of the PDF gives a result for $\tau$ that is well centered on the input value as desired, but has a tail of spurious detections at high time lags.  These spurious detections are far fewer when one uses the median to determine $\tau$, but the median distribution is not as well centred about the input value as the peak distribution. The desirable behavior would be a distribution that is centered on zero as the peak is, but without the high $\Delta\tau$ anomalies. To this end, we impose a restriction on a good recovery, requiring that the measurement determined by the peak and that determined by the median be consistent within a certain threshold, to be discussed in \Cref{determining_cut_size}. Since the peak measurement is best centered around $\Delta \tau =0$, we use this as the measure of $\tau$, with the proximity to the median measurement used as a filter to remove the spurious peak results that exist at a high $\Delta\tau$.

Requiring the peak and median measurements to agree within 100 days reduces the prominence of the artifacts present in both JAVELIN and ICCF at high \dt\ while retaining most of the accurate lag recoveries (see top panel of \Cref{fig:delta_tau_all_cuts}). In an attempt to mitigate the offset still present after applying this cut, we enforce another cut similar to that implemented in the top panel of \Cref{fig:delta_tau_all_cuts}, however, this time we only accept lags for which JAVELIN and ICCF agree.  Both methods should return the same lag for a reliable recovery, therefore we enforce that they must agree within a certain margin, the size of which we optimise in \Cref{determining_cut_size}.\footnote{ We choose to make these cuts absolute, rather than relative to the time lag, as the confidence that we have in a measurement relies upon how close the measurements are -- we have intrinsically more confidence in measurements that are 100 days apart than 300 days apart regardless of the underlying lag. An attempt was made to utilise relative cuts but this either made the cuts unreasonably small for short lags or unreasonably high for large lags. A relative cut may be more appropriate for samples where the range of time lags is smaller than that of the \ion{C}{iv} lags considered here.}  This cut further improves the accuracy of the recoveries, removing almost all of the remaining outliers at high \dt. These two cuts remove 26\% and 41\%  of realisations respectively, with 49\% removed with both cuts applied. Without these cuts the average bias is \dt=75 days for JAVELIN and \dt=141 days for ICCF.  After applying these first two cuts, those average biases are reduced to \dt=24 days and \dt=48 days respectively.  We note that this bias arises because of the skew in the distribution -- the median bias for JAVELIN is never over 6 days, even without cuts.  After cuts it is reduced to 2 days.  (ICCF starts at a median bias of 46 days without cuts, which reduces to  18 days after cuts).

\begin{figure}
    \centering
    \includegraphics[scale=0.58]{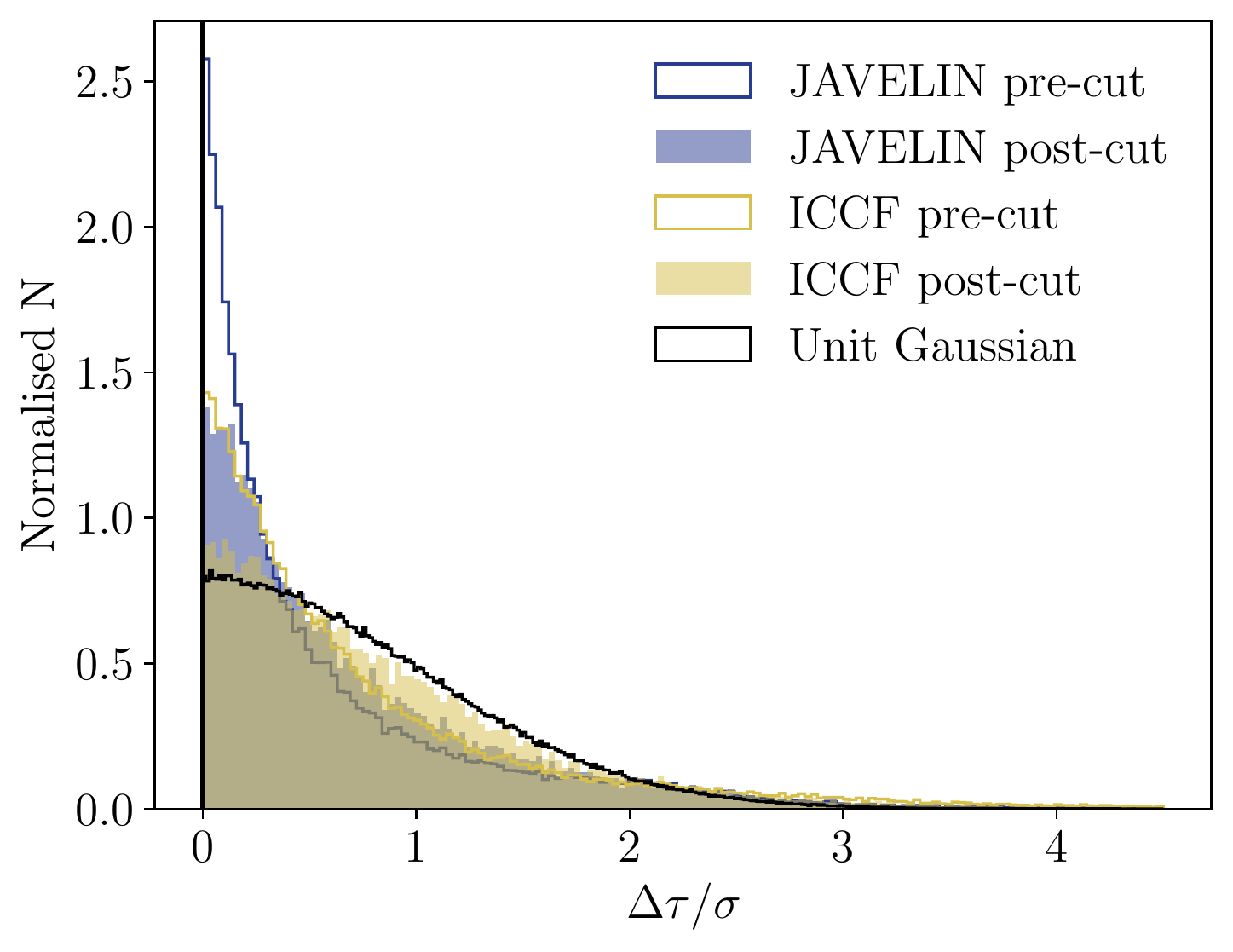}
    \caption{Comparison of $\Delta\sigma$ distributions for the mean absolute deviation for both JAVELIN and ICCF to a unit Gaussian.  Before any error cuts, both JAVELIN and ICCF have an overabundance of simulations with low $\Delta\sigma$, indicating that there are many simulations where the uncertainties are being overestimated by these methods. In light of this we apply an error cut of 80 days (exact size of cut to be determined in \Cref{determining_cut_size}). The results of applying this cut show that ICCF has uncertainties which closely match the unit Gaussian, while JAVELIN still overestimates the uncertainties.}
    \label{fig:delta_sigma_all_cuts}
\end{figure}

A summary of both the mean and median offsets at each cut level for both JAVELIN and ICCF, as well as the proportion of the full sample that pass those cuts, can be found in \Cref{delta_tau_summary_table_1}.

\begin{table*}
\centering
\caption{ The effect of cuts on the acceptance fraction and \dt\  offsets. Note that the mean and median offset are measured in days, with each consecutive cut improving the positive offset in both the mean and the median.}
\begin{tabular}{lccccc}
                &             & \multicolumn{2}{c}{Mean \dt}                         & \multicolumn{2}{c}{Median \dt}                            \\ \hline
After           & \% accepted & JAVELIN & ICCF & JAVELIN & ICCF \\ \hline
No cuts         & 100         & 75         & 142               & 6                         & 46                        \\ \hline
Cut 1           & 76          & 41         & 82                & 4                         & 33                        \\ \hline
Cut 2           & 59          & 52         & 59                & 3                        & 16                        \\ \hline
Cut 1 and Cut 2 & 51          & 24         & 48                & 2                        & 18                       
\end{tabular}
\label{delta_tau_summary_table_1}
\end{table*}

\subsection{Determining measurement uncertainties}\label{sect:uncertainties}
In addition to understanding the optimal way to measure time lags from PDFs, we also need to extract the most representative uncertainties. To do this we considered three methods to compute the uncertainty: the mean absolute deviation (Mean AD), the median absolute deviation (Median AD), and the area that encloses 34\% ($1\sigma$) of the probability on each side of the preferred lag (Count Out) as is used by default in JAVELIN. To assess the performance of each of these uncertainty measures, we used the relative distance of each simulation from its true lag or $\Delta\sigma=\frac{\Delta\tau}{\sigma}$ where $\sigma$ is the magnitude of the measurement uncertainty. Using the peak as the lag measure, we can test the comparative distributions for each of the methods for both JAVELIN and ICCF to determine the optimal measure and whether any error cuts are needed. We considered the absolute value of $\Delta\sigma$ as the sign correlates to the \dt\ value, therefore any biases in the sign were already shown in \Cref{fig:delta_tau_all_cuts}. We find that the count out method seemed to be the most representative, however, due to its asymmetric nature, it often displayed other undesirable behaviour, such as having uneven upper and lower uncertainties (e.g. -50,+500). Of the remaining two methods, the median absolute deviation closely matched a unitary Gaussian, as should be the case, however, it displayed an oversupply of higher $\Delta\sigma$, meaning that it often underestimated the size of the uncertainties. Given the option of a method that tends to overestimate uncertainties versus underestimate them, an overestimation is preferred as overestimated uncertainties may take into account unknown systematic errors and can be mitigated using quality cuts as was done in \Cref{sect:quality}.  Due to this we will conduct the remainder of this analysis using the mean absolute deviation as our uncertainty measure.

\begin{figure*}
    \centering
    \includegraphics[scale=0.68]{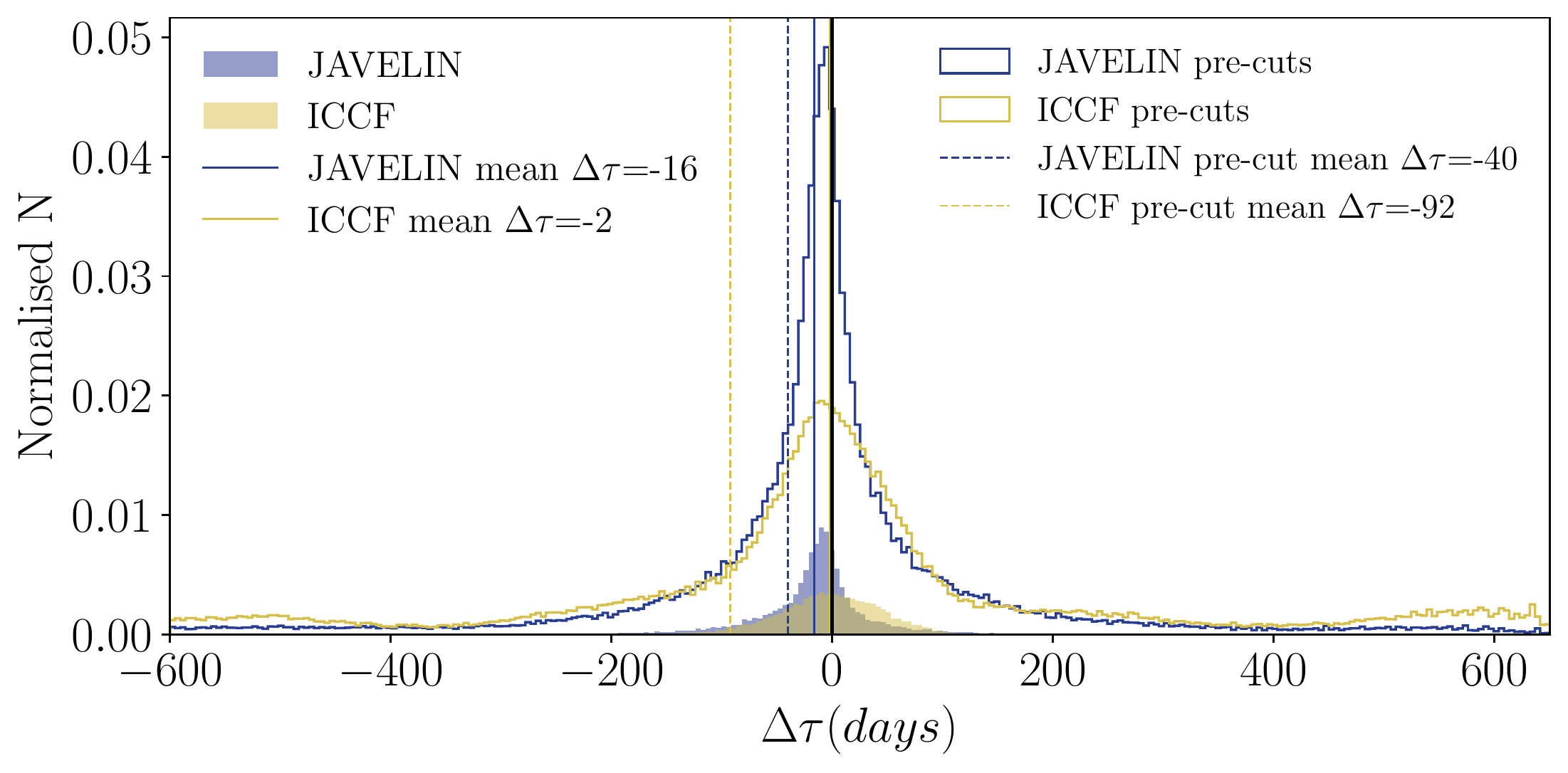}
    \caption{Distribution of $\Delta\tau$ using a prior centered around zero with the same cuts and processing as the bottom panel on \Cref{fig:delta_tau_all_cuts}. Note that the mean offset for the JAVELIN recoveries is consistent in magnitude with the that measured using only the positive prior range (consistent with 2/3 of the prior range now being below the expected lag, whereas previously 2/3 was above). The mean ICCF offset is closer to zero than that of JAVELIN, however it has a significantly larger dispersion.}
    \label{fig:neg_dtau}
\end{figure*}
Given that the confidence we have in a measurement is intrinsically tied to the uncertainty attached to that measurement, it is logical to place a cut on the absolute size of the error on each lag recovery. To show the effectiveness of this cut, we compare the normalised uncertainties for the mean absolute deviation before and after this cut. In \Cref{fig:delta_sigma_all_cuts} we have applied an 80 day uncertainty cut and can observe the effect this cut has on the behavior of the sample. Both JAVELIN and ICCF become much less peaked towards small $\Delta\sigma$, with ICCF in particular being very close to Gaussian. The slight peaking in the JAVELIN population indicates that the uncertainties are in general being overestimated.

With an extra cut being applied to the data, it is important to consider the effect it has on the \dt distribution. In \Cref{fig:delta_tau_all_cuts} we can see that this uncertainty cut has removed many sources from the central region of the distribution. This last cut further decreases the mean offsets; from 24 to 14 days for JAVELIN, and from 48 to 21 days for ICCF. This is still a positive offset, meaning that it may introduce a bias into future measurements such as black hole masses and the \textit{R-L} relation that would need to be corrected for. A possible solution to this would be making the prior range symmetric around 0, as opposed to using on the physical positive regime. However, as displayed in \Cref{fig:neg_dtau} there is a negligible difference between the bias these two methods exhibit besides a sign swap, with the negative baseline inclusion giving a negative bias of 16 days after the same three cuts are applied.  This indicates that the proportion of the prior range that exists above the `correct' answer versus below has a strong impact on the average offset. We could make the prior symmetric around the expected lag which would be more likely to give a minimal offset, but would also be strongly biasing ourselves toward what we expect. Instead we can use these simulations and modelling to account for biases that we find. Now that we better understand the behaviour of the simulation sample and how to extract the most representative lags, we can move to construct a statistically meaningful set of quality criteria through which to assign quality ratings to the observed \ion{C}{iv} sample.

\section{Criteria to establish lag measurements }\label{selection}
The purpose of studying the cuts discussed in \Cref{lag_rec} is to assign a quality rating to each recovery to encapsulate the reliability of each recovery. Each successful recovery will be given a ration of gold,silver or bronze with gold being the most reliable lags. In order to assess the efficacy of the quality cuts, we investigated the number of sources that pass the quality cuts discussed in \Cref{sect:quality} but recover the lag incorrectly. After these cuts approximately 15\% of the remaining measured lags produced by JAVELIN and ICCF didn't satisfy the following criteria:
\begin{itemize}
    \item $|\tau_{\text{sim}}-\tau_{\text{true}}| = $|\dt| < 80 days
    \item $\frac{\Delta\tau}{\sigma} = \Delta\sigma$ < 3
\end{itemize}
These false detections show that recovering incorrect lags remains possible even after applying the cuts. To mitigate this effect there are two courses of action: more stringent cuts to reduce the false detection rate, and a source-by-source test to measure this effect. Since we can individually simulate sources, a source by source false positive test will be implemented that considers the probability of recovering the lag that was recovered in each `real' source by chance. 

For any lag measured from a real lightcurve, we define the false positive rate (FPR) as the fraction of simulations, drawn from across the full range of simulated input lags for that source, that erroneously present as the measured lag with high confidence. This quantity will remove sources that include a systematic error in the time lag signal, for example a signal that arises from aliasing due to the observing cadence.

\begin{figure}
    \centering
    \includegraphics[scale=0.61]{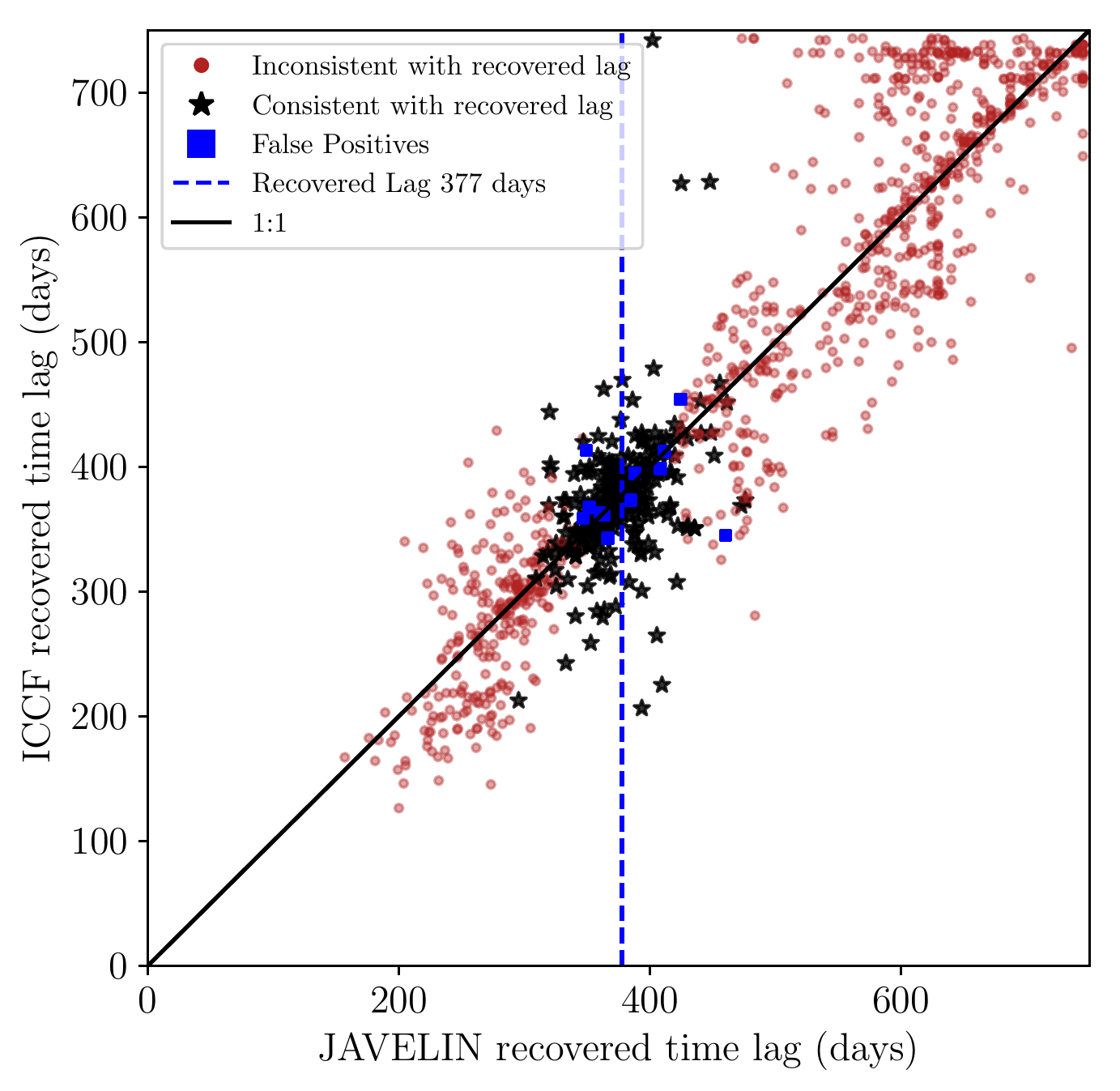}
    \caption{Graphical representation of the false positive test conducted on randomly selected \ion{C}{iv} source. The points on the plot represent all of the points that pass the quality cuts. We choose only the points that are consistent with the recovered lag in the data at a $1\sigma$ level, these are shown in black. The black stars show the points with an input lag consistent with the recovered lag. On the other hand the blue points show the simulations where the recovered lag is consistent with the lag recovered in the real data when the input to that simulation more than $3\sigma$ away from the point. From these we define our false positive rate as the number of blue points as a percentage of the number of blue and black combined.}
    \label{fig:false_positives}
\end{figure}

\Cref{fig:false_positives} graphically shows how we compute the FPR. The black points show the simulations where the recovered lag is within $1\sigma$ of the lag that was recovered in the observed lightcurve. Of these, we find the instances where the recovered lag disagrees with its input lag beyond a $3\sigma$ level, these are presented in blue. This is important as, assuming Gaussian errors, we would only expect 1\% to be inconsistent at this level. In the case pictured, the blue points make up $\sim 5\%$ of the non-red points, this means that having recovered a lag of $\sim 400$ days there is a $\sim 5\%$ chance that your recovery passes the first two quality cuts but is inconsistent with the physical lag in the lightcurves.
In addition to this test, we can require high quality recoveries pass more stringent cuts compared to those discussed in \Cref{lag_rec}.  We test the bulk behavior of the simulated sample with respect to these cuts, however, a cut on the source by source FPR will also be applied for each quality level. 

\subsection{Determining cut sizes}\label{determining_cut_size}
In Sections~\ref{sect:quality} and~\ref{sect:uncertainties}, we found that introducing three data quality cuts improved the accuracy of lag recoveries as well as removing the artifacts at high \dt. For that initial proof of concept we used a threshold of 100 days for the first two cuts (peak agrees with median, and JAVELIN agrees with ICCF) and 80 days for the third cut (uncertainty). However, to determine the optimal size of these cuts we now consider the impact of the cut size on four lag-quality measures:
\begin{enumerate}
    \item the average offset from the true lag (mean $\Delta\tau$);
    \item the median offset from the true lag (median $\Delta\tau$);
    \item the false detection rate (\Cref{selection}); and
    \item the proportion of the simulations that pass the cuts.
\end{enumerate}
We varied the three cuts from 30 to 200 days, and examined the four lag-quality measures in each case. The ideal set of cuts would have low offsets and false detection rates, with a high acceptance fraction. In practice as we loosen the cuts the acceptance fraction increases but the quality declines.  We use this to define a set of quantitative quality cuts.

First we considered the effect of each of the cuts separately on each of the measurable quantities described previously. We find that the cut on the agreement between JAVELIN and ICCF produces the same average result for mean offset, acceptance fraction and false detection rate regardless of the size of the cut (between 30-200 days). In light of this, we set 100 days as the maximum acceptable difference between measurements made by JAVELIN and ICCF.
\begin{figure*}
    \centering
    \hspace*{-0.7cm}\includegraphics[scale=0.51]{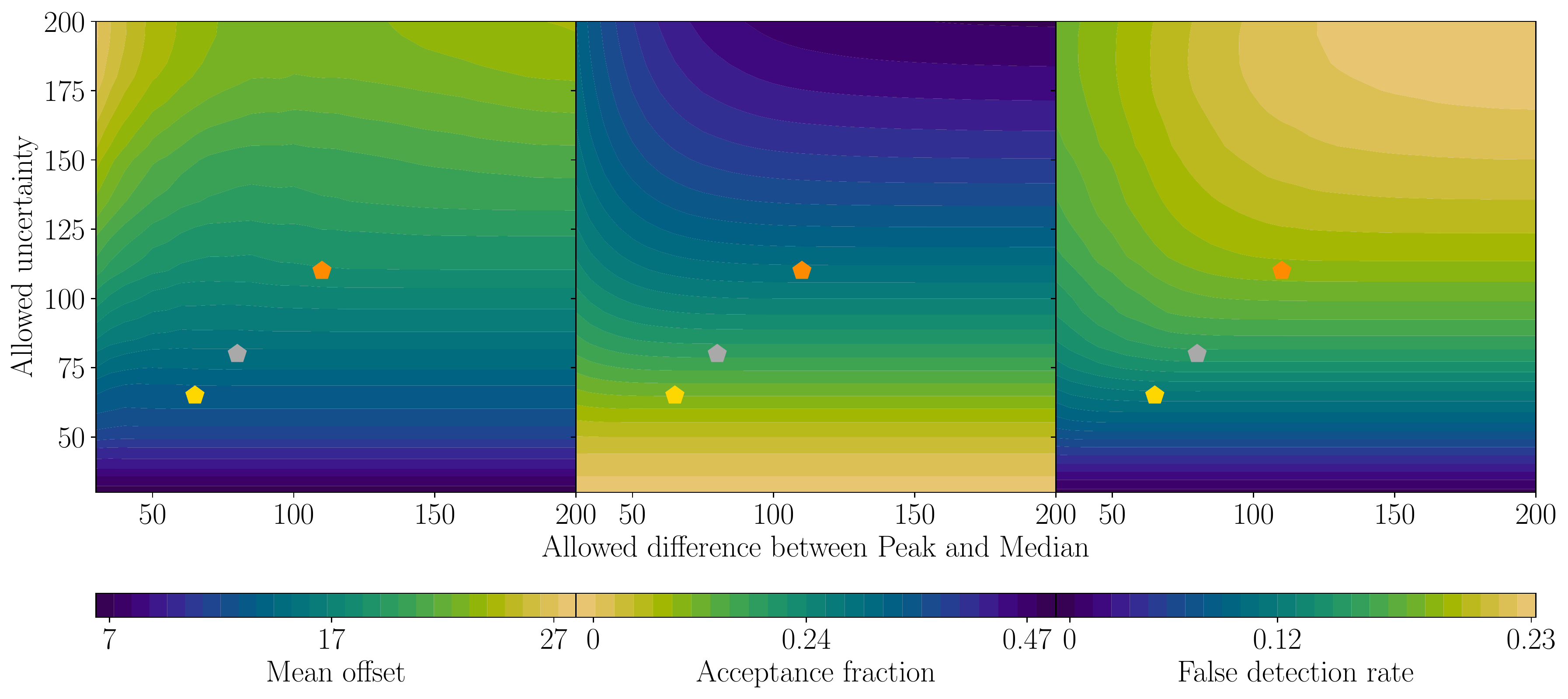}
    \caption{Contours showing the characteristics of the full simulation sample after cuts. We can see that the uncertainty cut is the main driver for all three of the measured quantities, with the difference between the peak and median measurements having more of an impact when the uncertainty cut is relaxed. Note that the bluer areas denote the desired outcome with yellower areas being less desirable. The decided upon cuts are denoted by the gold, silver and bronze points.}
    \label{accept_cont}
\end{figure*}

With this cut made, we can visualise the effect of the other two cuts on the mean offset, acceptance fraction and false detection rate as shown in \Cref{accept_cont}. It is obvious that the cut on uncertainty size is the main driver for low mean offset and false detection rate, however, the difference between peak and median offset also makes an impact in certain regimes. Using this information, we can choose thresholds in both of these cuts that provide differing qualities of recovery while still giving reliable recoveries across the board. The chosen cuts are shown in \Cref{Quality_ratings}.  

\begin{table*}

\begin{tabular}{lcccccccc}
        & \multicolumn{3}{c}{Cut Size (days)}             &&  \multicolumn{4}{c}{Quality Measure (days or \%)} \\ \cline{2-4}\cline{6-9}\\
Quality & JAVELIN-ICCF & Peak-Median & Uncertainty && Mean $\Delta\tau$ & Median $\Delta\tau$ & False detection \% & \% accepted \\ \hline
Gold    & 100          & 65          & 65          && 13                & 6                   & 12              & 12\%          \\ \hline
Silver  & 100          & 80          & 80          && 15                & 6                   & 15              & 19\%          \\ \hline
Bronze  & 100          & 110         & 110         && 18                & 6                   & 19              & 29\%          
\end{tabular}
\caption{Statistics for the samples that satisfy each of the different quality criteria. The median offset does not change significantly between cuts; however, the average offset does. This indicates that there is a systematic offset that is present at all cut levels, with the cuts mostly removing outliers that skew the average to higher offsets. Also note that the percentage accepted includes those in the quality tier above (e.g.\ Bronze includes those that pass Silver and Gold). }
\label{Quality_ratings}
\end{table*}

\section{$\textit{R-L}$ relation analysis }\label{rldata}
One of the important products that is generated through analysing an RM data sample is the Radius-Luminosity ($\textit{R-L}$) relation. Given that we have access to a large number of representative simulations, we have the opportunity to test the accuracy with which we can measure the $\textit{R-L}$ relation given that the input $\textit{R-L}$ relation for the simulations is known. To test the effectiveness of the quality criteria, we can fit the $\textit{R-L}$ relation using only sources that fit into each quality criterion. With the large number of simulations we have generated we can do these fits many times choosing a different subset of measurements each time. This will allow us to not only observe the overall quality of the fits but also the effect that the cuts have on the $\textit{R-L}$ relation fits. For each iteration of this test, a sample of recovered lags was chosen that contained only one realisation of any one source, with the fit being repeated 1000 times to determine the reliability of the fits. 

It is important to note that our sources cover a fairly small range of radii and luminosities. Due to this we have included literature measurements to anchor the R-L relation fits at the low and high luminosity ends. Combining data sets in this way is standard when generating the \ion{C}{iv} $\textit{R-L}$ relation.  The extra sources cover a luminosity range of 43.6 erg s$^{-1}\lesssim \log\lambda L_{\lambda} \lesssim$ 47.2 erg s$^{-1}$ and time delay range of 4 days $\lesssim \tau_{\rm RF}\lesssim460$ days.  Detailed information on these sources is shown in \Cref{tab:existingLags} and visualised relative to our data in \Cref{fig:RL_single}.

\begin{figure*}
    \centering
    \hspace*{-0.7cm}\includegraphics[scale=0.61]{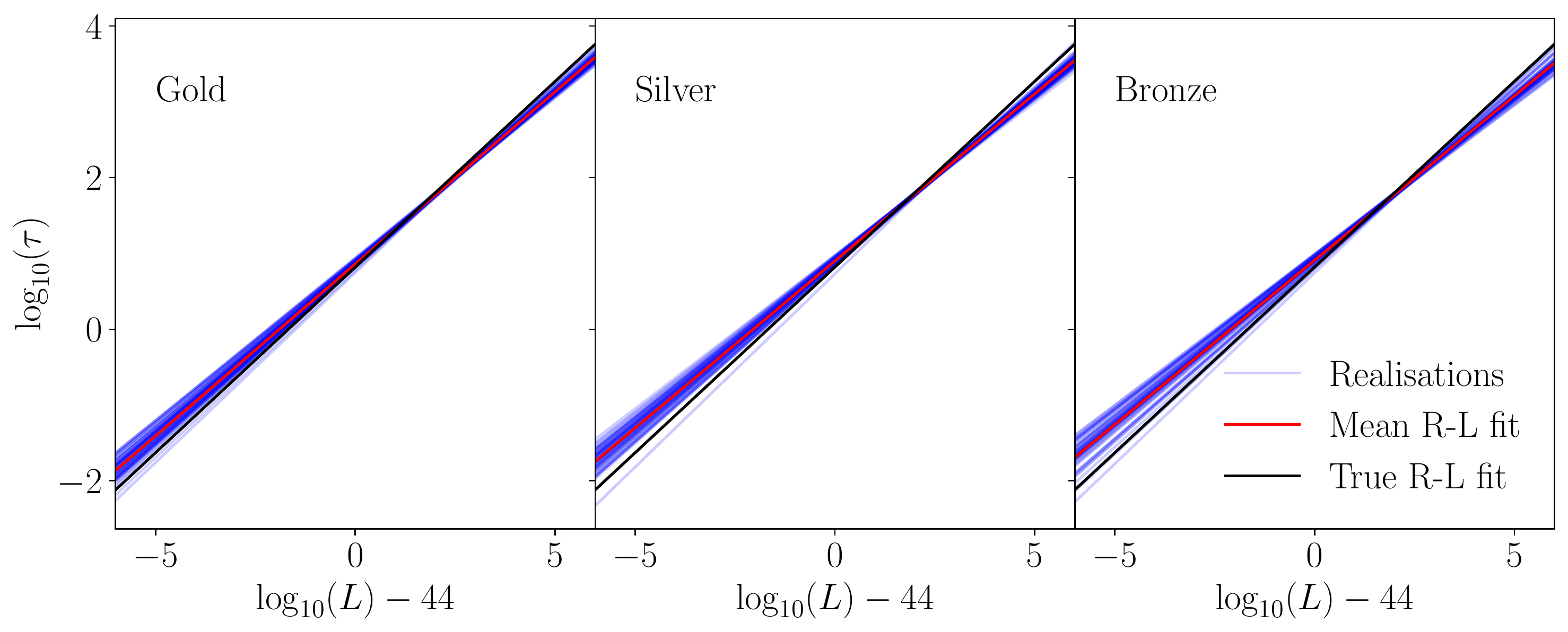}
    \caption{Best fit $\textit{R-L}$ relations for a sample of 50 iterations, computed using a subsample of sources from each cut group. The cuts have an obvious effect on the quality of the $\textit{R-L}$ relation recovery with the gold standard fits having very little variation from true fit}
    \label{fig:RLfits}
\end{figure*}

\begin{figure*}
    \centering
    \hspace*{-0.7cm}\includegraphics[scale=0.61]{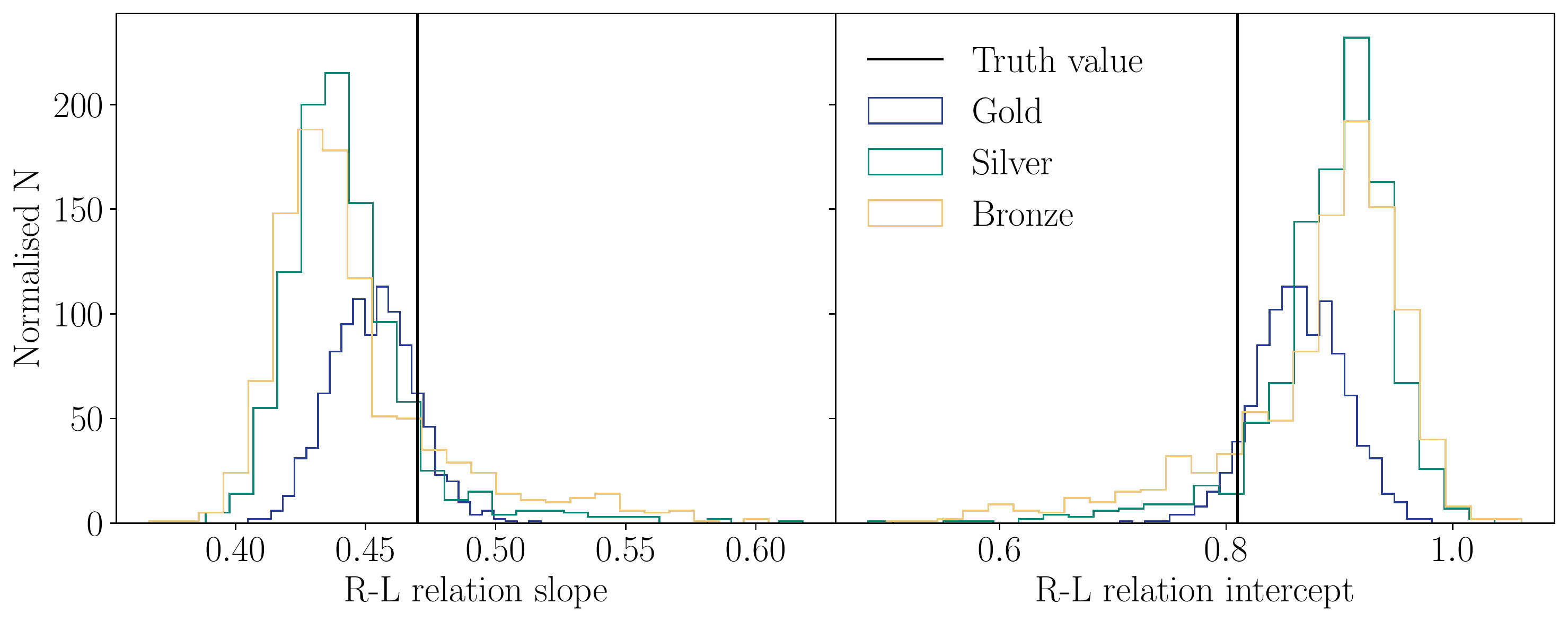}
    \caption{Best fit $\textit{R-L}$ relation slopes ($\frac{\log_{10}(\text{days})}{\log_{10}(\text{erg}\cdot\text{ s}^{-1})}$) and intercepts ($\log_{10}(\text{days})$) computed using a subsample of sources from each cut group. These histograms contain 1000 iterations of the fitting procedure. The cuts have an obvious effect on the quality of the $\textit{R-L}$ relation recovery with the gold standard fits being centered around the truth value for the slope and closest of the three groups for the intercept. Next most reliable is silver and then bronze as expected. All three distributions are offset from the truth value in regards to the intercept, this is likely due to the positive offset seen in the simulation results and may be compensated for if well understood. Note that the mean uncertainties in the intercept measurements are approximately 0.034 for all cuts, with the average uncertainty in the intercept being 0.070, 0.077 and 0.086  for the gold, silver and bronze level cuts respectively.}
    \label{fig:RLfits_goodness}
\end{figure*}
The effect that the cuts have on the $\textit{R-L}$ relation fits is shown in \Cref{fig:RLfits} and \Cref{fig:RLfits_goodness}. The gold standard cuts provide the most accurate fit on average, and also display the least variance in the fits with a mean slope of 0.454 and a standard deviation in slope of 0.016. In contrast, the silver and bronze fits have slopes and deviations of $0.445\pm 0.027$ and $0.447\pm 0.034$ respectively. The average fits for each of the three quality tiers are within $1\sigma$ of the true value for the $\textit{R-L}$ relation, suggesting that they are all reliable, with the gold standard simply being the most well constrained.  

All three of the quality levels do display slight offsets in both the slope and the intercept from the input value (\Cref{fig:RLfits}). Given the small residual offset we found in the lag recoveries, this is to be expected.  Even though the offsets are less than the 1$\sigma$ uncertainties, we would like to take them into account and derive an even more accurate relation. The ultimate source of the bias comes from the imperfect or incomplete observational data (e.g. sampling limitations, survey window functions, malmquist biases).  Some of those are irreducible systematic uncertainties, but with a thorough simulation suite some of these can be accounted for \citep[as is done in many astrophysics applications such as supernova cosmology, e.g. ][]{Kessler2019}. 
 In \Cref{bias_correction} we show that simply adding the mean time-delay offset ($\Delta\tau$) between the simulated time delays and the recovered time delays to each AGN data point ($\tau$) improves the recovery of the $\textit{R-L}$ relation. We defer a more thorough examination of bias-correction methods to a future paper.  Even without any further correction, our recovery of the $\textit{R-L}$ relation is accurate to within $1\sigma$, so it is already useful for many astrophysical questions (such as the relative size of black holes at different epochs), but the bias will be critical to address for applications such as using AGN as standard candles for cosmology.

\section{Summary}\label{summary}
In this paper we have developed an extensive set of simulations that can be used to model individual sources based on their exact variability, cadence and other observable traits. We use this powerful tool to quantify the lag recovery of each AGN in our sample rather than derive summary results from a observable distributions of parameters.
 
Using these simulations we investigated how best to extract a time lag from the posterior lag distribution produced by JAVELIN and ICCF. We found that the peak of this distribution is the best measure of the time lag and the mean absolute deviation is the best measure of the uncertainty. From there we developed a set of quality cuts to show which of the recoveries exhibit the characteristics of a reliable time lag. We use cuts based on the deviation between ICCF and JAVELIN recoveries, the deviation between the peak and median of the underlying PDF for each recovery, and on the absolute uncertainty of the lag. We also designed a false positive test that quantifies the likelihood that the recovered lag has been measured in an incorrect location. Using these cuts, we implemented a gold, silver and bronze rating system, and used these ratings to test the quality of the resulting Radius-Luminosity relation. All of these quality levels produced $\textit{R-L}$ relations that were correct within a $1\sigma$ confidence level, with the gold sample producing the least variance in fits.

In future work we aim to make improvements to several aspects of our analysis:
\begin{itemize}
     \item Where in this paper we estimated the spectroscopic calibration uncertainties using the method described in \citet{Hoormann2019}, we are developing a new empirical model of estimating the calibration uncertainties based on the F-star spectra for the upcoming OzDES papers. This has been implemented for MgII measurements \citep{Zhefu2021} and we will apply it to the \ion{C}{iv} region in the future.\newline
    \item The photometry used in this analysis was measured in the $g-$band, however, we also have $r-$ and $i-$band magnitudes. These are currently used only in our spectrophotometric calibration but it is possible that they could be used in constraining time lags as well. Using JAVELIN it is possible to cross-correlate multiple photometric bands as well as the spectroscopic lightcurve, in theory this would provide better constraints. The slight time-delay between the photometric bands due to the continuum emission for the host black hole's accretion disk \citep{Mudd2018}, would provide slightly different time domain information and may reduce effects such as aliasing. This method would be computationally expensive and thus was not explored here but is a consideration moving forward.
\end{itemize}

 With a well understood set of criteria to help understand our confidence in measurements made on the full DES/OzDES sample, we now have a strong frame work on which to build the bulk analysis of the remainder of the data set (Penton et al. in prep, Malik et al. in prep, Yu et al. in prep). 
 
{\black
\section*{Acknowledgements}
This research was supported in part by the Australian Government through the Australian Research Council Laureate Fellowship funding scheme (project FL180100168). 

AP and UM are supported by the Australian Government Research Training Program (RTP) Scholarship.

PM and ZY were supported in part by the United States National Science Foundation under Grant No. 161553.

Funding for the DES Projects has been provided by the U.S. Department of Energy, the U.S. National Science Foundation, the Ministry of Science and Education of Spain, the Science and Technology Facilities Council of the United Kingdom, the Higher Education Funding Council for England, the National Center for Supercomputing Applications at the University of Illinois at Urbana-Champaign, the Kavli Institute of Cosmological Physics at the University of Chicago, the Center for Cosmology and Astro-Particle Physics at the Ohio State University, the Mitchell Institute for Fundamental Physics and Astronomy at Texas A\&M University, Financiadora de Estudos e Projetos, Funda{\c c}{\~a}o Carlos Chagas Filho de Amparo {\`a} Pesquisa do Estado do Rio de Janeiro, Conselho Nacional de Desenvolvimento Cient{\'i}fico e Tecnol{\'o}gico and the Minist{\'e}rio da Ci{\^e}ncia, Tecnologia e Inova{\c c}{\~a}o, the Deutsche Forschungsgemeinschaft and the Collaborating Institutions in the Dark Energy Survey. 

The Collaborating Institutions are Argonne National Laboratory, the University of California at Santa Cruz, the University of Cambridge, Centro de Investigaciones Energ{\'e}ticas, Medioambientales y Tecnol{\'o}gicas-Madrid, the University of Chicago, University College London, the DES-Brazil Consortium, the University of Edinburgh, the Eidgen{\"o}ssische Technische Hochschule (ETH) Z{\"u}rich, Fermi National Accelerator Laboratory, the University of Illinois at Urbana-Champaign, the Institut de Ci{\`e}ncies de l'Espai (IEEC/CSIC), the Institut de F{\'i}sica d'Altes Energies, Lawrence Berkeley National Laboratory, the Ludwig-Maximilians Universit{\"a}t M{\"u}nchen and the associated Excellence Cluster Universe, the University of Michigan, NFS's NOIRLab, the University of Nottingham, The Ohio State University, the University of Pennsylvania, the University of Portsmouth, SLAC National Accelerator Laboratory, Stanford University, the University of Sussex, Texas A\&M University, and the OzDES Membership Consortium.
Based in part on observations at Cerro Tololo Inter-American Observatory at NSF's NOIRLab (NOIRLab Prop. ID 2012B-0001; PI: J. Frieman), which is managed by the Association of Universities for Research in Astronomy (AURA) under a cooperative agreement with the National Science Foundation.

The DES data management system is supported by the National Science Foundation under Grant Numbers AST-1138766 and AST-1536171.
The DES participants from Spanish institutions are partially supported by MICINN under grants ESP2017-89838, PGC2018-094773, PGC2018-102021, SEV-2016-0588, SEV-2016-0597, and MDM-2015-0509, some of which include ERDF funds from the European Union. IFAE is partially funded by the CERCA program of the Generalitat de Catalunya.
Research leading to these results has received funding from the European Research
Council under the European Union's Seventh Framework Program (FP7/2007-2013) including ERC grant agreements 240672, 291329, and 306478.
We  acknowledge support from the Brazilian Instituto Nacional de Ci\^encia
e Tecnologia (INCT) do e-Universo (CNPq grant 465376/2014-2).

This manuscript has been authored by Fermi Research Alliance, LLC under Contract No. DE-AC02-07CH11359 with the U.S. Department of Energy, Office of Science, Office of High Energy Physics.

\section*{Data Availability}
The data underlying this article are available in DES data release 2, at \url{https://des.ncsa.illinois.edu/releases/dr2} and the OzDES data release 2 at \url{https://datacentral.org.au/services/download/}. The datasets were derived from sources in the public domain \cite{2021DES,2020OzDES}.




\bibliographystyle{mnras}
\bibliography{refs} 
\vspace{0.4cm}

$^{1}$ School of Mathematics and Physics, University of Queensland,  Brisbane, QLD 4072, Australia\\
$^{2}$ The Research School of Astronomy and Astrophysics, Australian National University, ACT 2601, Australia\\
$^{3}$ Center for Cosmology and Astro-Particle Physics, The Ohio State University, Columbus, OH 43210, USA\\
$^{4}$ Department of Astronomy, The Ohio State University, Columbus, OH 43210, USA\\
$^{5}$ Radcliffe Institute for Advanced Study, Harvard University, Cambridge, MA 02138\\
$^{6}$ Centre for Gravitational Astrophysics, College of Science, The Australian National University, ACT 2601, Australia\\
$^{7}$ Departamento de F\'isica Matem\'atica, Instituto de F\'isica, Universidade de S\~ao Paulo, CP 66318, S\~ao Paulo, SP, 05314-970, Brazil\\
$^{8}$ Laborat\'orio Interinstitucional de e-Astronomia - LIneA, Rua Gal. Jos\'e Cristino 77, Rio de Janeiro, RJ - 20921-400, Brazil\\
$^{9}$ Fermi National Accelerator Laboratory, P. O. Box 500, Batavia, IL 60510, USA\\
$^{10}$ Centro de Investigaciones Energ\'eticas, Medioambientales y Tecnol\'ogicas (CIEMAT), Madrid, Spain\\
$^{11}$ Institute of Cosmology and Gravitation, University of Portsmouth, Portsmouth, PO1 3FX, UK\\
$^{12}$ CNRS, UMR 7095, Institut d'Astrophysique de Paris, F-75014, Paris, France\\
$^{13}$ Sorbonne Universit\'es, UPMC Univ Paris 06, UMR 7095, Institut d'Astrophysique de Paris, F-75014, Paris, France\\
$^{14}$ Department of Physics and Astronomy, Pevensey Building, University of Sussex, Brighton, BN1 9QH, UK\\
$^{15}$ Department of Physics \& Astronomy, University College London, Gower Street, London, WC1E 6BT, UK\\
$^{16}$ Instituto de Astrofisica de Canarias, E-38205 La Laguna, Tenerife, Spain\\
$^{17}$ INAF - Astronomical Observatory of Trieste, I-34143, Trieste, Italy\\
$^{18}$ Center for Astrophysical Surveys, National Center for Supercomputing Applications, 1205 West Clark St., Urbana, IL 61801, USA\\
$^{19}$ Department of Astronomy, University of Illinois at Urbana-Champaign, 1002 W. Green Street, Urbana, IL 61801, USA\\
$^{20}$ Institut de F\'{\i}sica d'Altes Energies (IFAE), The Barcelona Institute of Science and Technology, Campus UAB, 08193 Bellaterra (Barcelona) Spain\\
$^{21}$ INAF-Osservatorio Astronomico di Trieste, via G. B. Tiepolo 11, I-34143 Trieste, Italy\\
$^{22}$ Institute for Fundamental Physics of the Universe, Via Beirut 2, 34014 Trieste, Italy\\
$^{23}$ Observat\'orio Nacional, Rua Gal. Jos\'e Cristino 77, Rio de Janeiro, RJ - 20921-400, Brazil\\
$^{24}$ Department of Physics, University of Michigan, Ann Arbor, MI 48109, USA\\
$^{25}$ Department of Astronomy/Steward Observatory, University of Arizona, 933 North Cherry Avenue, Tucson, AZ 85721-0065, USA\\
$^{26}$ Jet Propulsion Laboratory, California Institute of Technology, 4800 Oak Grove Dr., Pasadena, CA 91109, USA\\
$^{27}$ Santa Cruz Institute for Particle Physics, Santa Cruz, CA 95064, USA\\
$^{28}$ Institute of Theoretical Astrophysics, University of Oslo. P.O. Box 1029 Blindern, NO-0315 Oslo, Norway\\
$^{29}$ Institut d'Estudis Espacials de Catalunya (IEEC), 08034 Barcelona, Spain\\
$^{30}$ Institute of Space Sciences (ICE, CSIC),  Campus UAB, Carrer de Can Magrans, s/n,  08193 Barcelona, Spain\\
$^{31}$ Kavli Institute for Cosmological Physics, University of Chicago, Chicago, IL 60637, USA\\
$^{32}$ Instituto de Fisica Teorica UAM/CSIC, Universidad Autonoma de Madrid, 28049 Madrid, Spain\\
$^{33}$ Department of Astronomy, University of Michigan, Ann Arbor, MI 48109, USA\\
$^{34}$ Department of Physics, Stanford University, 382 Via Pueblo Mall, Stanford, CA 94305, USA\\
$^{35}$ Kavli Institute for Particle Astrophysics \& Cosmology, P. O. Box 2450, Stanford University, Stanford, CA 94305, USA\\
$^{36}$ SLAC National Accelerator Laboratory, Menlo Park, CA 94025, USA\\
$^{37}$ Department of Physics, The Ohio State University, Columbus, OH 43210, USA\\
$^{38}$ Center for Astrophysics $\vert$ Harvard \& Smithsonian, 60 Garden Street, Cambridge, MA 02138, USA\\
$^{39}$ Lawrence Berkeley National Laboratory, 1 Cyclotron Road, Berkeley, CA 94720, USA\\
$^{40}$ Australian Astronomical Optics, Macquarie University, North Ryde, NSW 2113, Australia\\
$^{41}$ Lowell Observatory, 1400 Mars Hill Rd, Flagstaff, AZ 86001, USA\\
$^{42}$ George P. and Cynthia Woods Mitchell Institute for Fundamental Physics and Astronomy, and Department of Physics and Astronomy, Texas A\&M University, College Station, TX 77843,  USA\\
$^{43}$ Instituci\'o Catalana de Recerca i Estudis Avan\c{c}ats, E-08010 Barcelona, Spain\\
$^{44}$ Physics Department, 2320 Chamberlin Hall, University of Wisconsin-Madison, 1150 University Avenue Madison, WI  53706-1390\\
$^{45}$ Universite Clermont Auvergne, CNRS/IN2P3, LPC, F-63000 Clermont-Ferrand, France\\
$^{46}$ Institute of Astronomy, University of Cambridge, Madingley Road, Cambridge CB3 0HA, UK\\
$^{47}$ Department of Astrophysical Sciences, Princeton University, Peyton Hall, Princeton, NJ 08544, USA\\
$^{48}$ Department of Physics, Duke University Durham, NC 27708, USA\\
$^{49}$ School of Physics and Astronomy, University of Southampton,  Southampton, SO17 1BJ, UK\\
$^{50}$ Computer Science and Mathematics Division, Oak Ridge National Laboratory, Oak Ridge, TN 37831\\
$^{51}$ McDonald Observatory, The University of Texas at Austin, Fort Davis, TX 79734\\
$^{52}$ Max Planck Institute for Extraterrestrial Physics, Giessenbachstrasse, 85748 Garching, Germany\\
$^{53}$ Universit\"ats-Sternwarte, Fakult\"at f\"ur Physik, Ludwig-Maximilians Universit\"at M\"unchen, Scheinerstr. 1, 81679 M\"unchen, Germany\\
}



\appendix
\section{Simulation Parameters}\label{sim_param}
\subsection{Variability time-scale, $\tau_D$, and amplitude, $SF_\infty$}

\citet{MacLeod2010} determined the following powe\textit{R-L}aw relationship between $\tau_D$ and $SF_\infty$ and the physical properties of AGN:
\begin{align}
\log_{10}(\alpha) = A_\alpha + B_\alpha \log_{10}\left(\frac{\lambda}{4000}\right) &+ C_\alpha(M_i + 23)\nonumber \\
&+D_\alpha \log_{10}\left(\frac{M_{\textrm{BH}}}{10^9 M_\odot}\right)
\end{align}
where $\alpha$ refers to $\tau_D$ (in days) or $SF_\infty$ (in mag), $\lambda$ (\AA) is the rest-frame continuum wavelength, $M_i$ is the absolute $i$-band magnitude of the source and $M_{\textrm{BH}}$ is the mass of the black hole (in solar masses). The coefficients are given in Table \ref{t:2}. The correlations were found after converting timescales to the rest-frame of the sources, therefore the estimated $\tau_D$ values were multiplied by ($1+z$) to create simulated light curves in the observed frame. 

\begin{table}
\centering
\caption{Coefficients of the power law relation defined for the damping timescale and structure function.}
\label{t:2}
\vspace{3mm}
\begin{tabular}{lll}
  & $\tau_D$ & $SF_\infty$   \\ 
\hline
$A$ & 2.4    & -0.51  \\
$B$ & 0.17   & -0.48  \\
$C$ & 0.03   & 0.13   \\
$D$ & 0.21   & 0.18   \\
\end{tabular}
\end{table}

The black hole mass that is often used in simulations of this type is that predicted from the following Gaussian distribution used by \citet{MacLeod2010}: 
\begin{align}\label{BHMass}
P(\log_{10} M_{\textrm{BH}} | M_i ) = \frac{1}{\sqrt[]{2 \pi \sigma^2_{M_{\textrm{BH}}}}} \exp \qty[-\frac{(\log_{10}M_{\textrm{BH}} - \log_{10}\overline{M}_{\textrm{BH}} )^2}{2 \sigma^2_{M_{\textrm{BH}}}}]
\end{align}
with mean $\log_{10}\overline{M}_{\textrm{BH}} = 2.0 - 0.27 M_i$ and standard deviation $\sigma_{M_{\textrm{BH}}} = 0.58 + 0.011 M_i$ which signifies the dispersion in the black hole population. \\
\\
In general, \Cref{BHMass} can be used to estimate the black hole masses for a large population of SMBH, However, it was found that, due to our survey target selection, the expected masses for our sources were tightly localised to much smaller regions than this distribution would predict.\\
\begin{figure}
    \centering
    \includegraphics[scale=0.6]{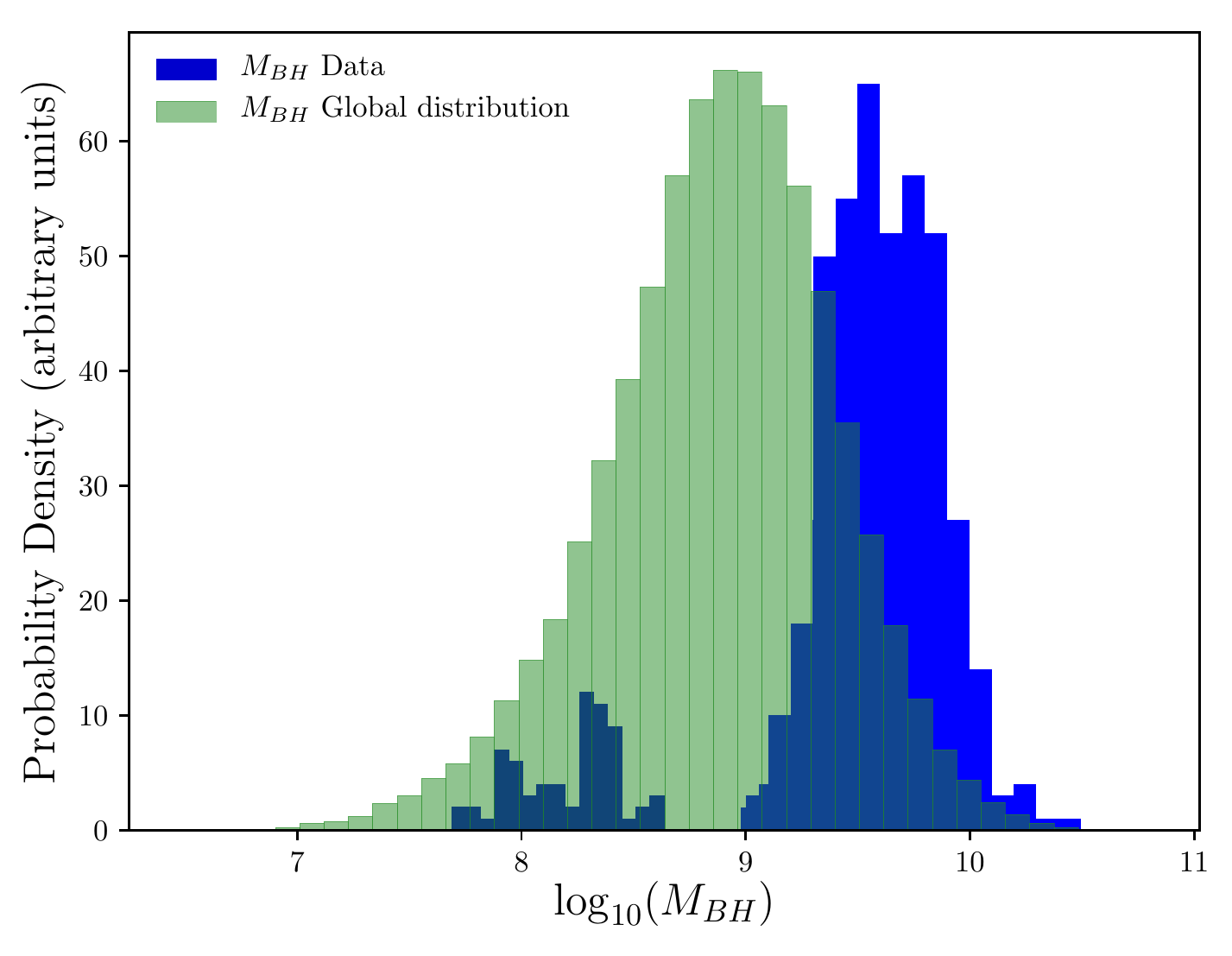}
    \caption{Histograms showing that our lag measurements primarily sample the high-mass end of the black hole mass distribution predicted by \Cref{BHMass}.  This is expected since we are primarily sensitive to long lags. The data in blue was obtained using the lag estimated from the $\textit{R-L}$ relation listed in \citet{Hoormann2019} and the line velocities from the RMS spectra for each source. These are then used to compute a mass using the virial relation (\Cref{virial}). The green distribution is the prediction of the black hole mass distribution using \Cref{BHMass} and the absolute $i-$band magnitudes for each source.}
    \label{fig:BHMdist}
\end{figure}
\Cref{fig:BHMdist} shows the difference in distributions between the masses for the OzDES sample as predicted by \Cref{BHMass} in comparison with the distribution of these same black hole masses estimated using the virial relation (\Cref{virial}). To use the virial relation we have used the $\textit{R-L}$ relation from \citet{Hoormann2019} to find an approximate radius of the BLR.  Then used the RMS spectrum for each source to find the velocity of the region. The two distinct groups of blue in \Cref{fig:BHMdist} represent the $H\beta$ sources ($\sim 10^8 M_\odot$) and the C~\textsc{iv} and Mg~\textsc{ii} sources ($\sim 10^9-10^{10.5} M_\odot$). The reason for the distinct groups is due to both survey design and astrophysical constraints. All of the nearby sources utilise the $H\beta$ line and are generally smaller/dimmer sources. This in general means that they host a lower mass SMBH. At a higher redshift we target much brighter objects, which generally house much larger SMBHs. There are also simply no quasars that house ($> 10^9 M_\odot$) SMBH at $z<0.6$ in our survey footprint. Due to these differences from the commonly accepted population distribution we conclude that the distributions in blue are much more likely to represent the physical characteristics of our sample. The BH mass that is used in our simulations is drawn randomly from these distributions, split into the appropriate emission line, as opposed to the global distribution from \cite{MacLeod2010} that is commonly used.

\subsection{Expected lag, $\tau$}\label{sec:lag}

We estimated the rest-frame lag for each source using published $\textit{R-L}$ relationships for each of the emission lines, which have the form:
\begin{equation}
\log_{10} (R) = K + \alpha \times \log_{10}(\lambda L_\lambda ) \label{eq:RL}
\end{equation}
where $R$ is the radius of the BLR in light-days (\textit{i.e.,}\ the lag, $\tau$, in days), $L_\lambda$ is the monochromatic continuum luminosity at wavelength $\lambda$\,(\AA) in erg\,s$^{-1}$\,\AA$^{-1}$, $K$ is the zero point for the relation, and $\alpha$ is the slope of the power law relationship. We use the $\textit{R-L}$ relation calibrated for C~\textsc{iv} from \citet{Hoormann2019} with coefficients of $K=-20.74 \pm 2.2$, $\alpha=0.47 \pm 0.04$ and $\lambda$\,(\AA)$=1350$. The simulated light curves were generated in the observed frame, so to generate the expected observe-frame lags for each source, the rest-frame lags were multiplied by ($1+z$).


\section{Radius-Luminosity relation fitting and corrections}\label{bias_correction}
\begin{figure*}
    \centering
    \hspace*{-0.7cm}\includegraphics[scale=0.61]{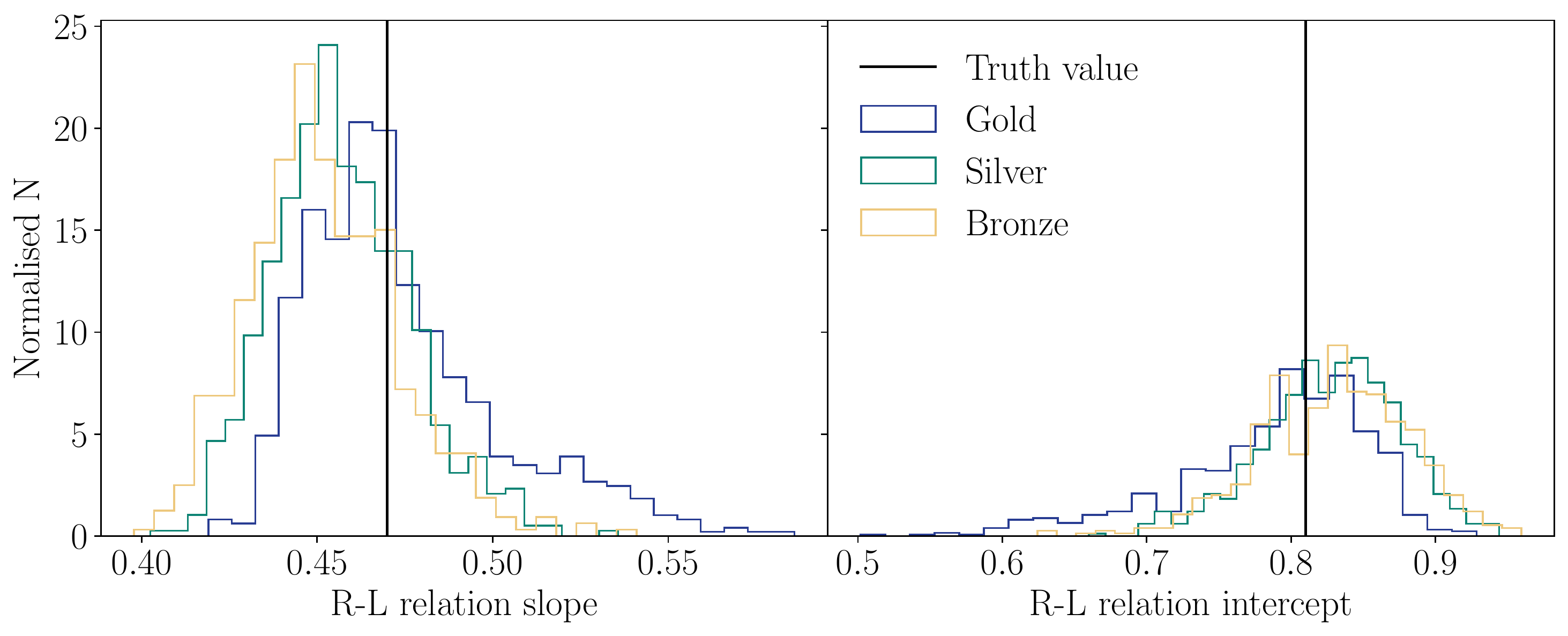}
    \caption{Best fit \textit{R-L} relation slopes ($\frac{\log_{10}(\text{days})}{\log_{10}(\text{erg}\cdot\text{ s}^{-1})}$) and intercepts ($\log_{10}(\text{days})$) computed using a subsample of sources from each cut group, with corrections for each based on the average offset shown in \Cref{Quality_ratings}. We can see that the biases seen in \Cref{fig:RLfits_goodness} are greatly reduced. This indicates that using these simulations to account for biases can be effective.}
    \label{fig:RL-corrected}
\end{figure*}

As mentioned in \Cref{rldata} we use some some previous \ion{C}{iv} lag measurements to help anchor the Radius-Luminosity relationship for our simulated data.  These are shown in \Cref{tab:existingLags}.  In order to assess whether the small residual bias in the \textit{R-L} relation can be removed, we take the average magnitude offset we see in each of our Gold, Silver, and Bronze samples, and add that offset to the results for each time delay in our simulated sample.  For example, the Gold sample had a mean offset $\Delta\tau=13$ days (\Cref{Quality_ratings}), so we add that to each recovered $\tau$ before fitting the \textit{R-L} relation.  The result is shown in \Cref{fig:RL-corrected}, in which an improved recovery of the relation is achieved.  

We note that this offset is not applied to previous data because that sample would have different statistical properties that would require its own simulation analysis.  A more sophisticated technique would be to apply a different offset for different subsets of the data, e.g. as a function of luminosity or lag.  We defer such explorations to future work.

\begin{figure}
    \centering
    \hspace*{-0.7cm}\includegraphics[scale=0.61]{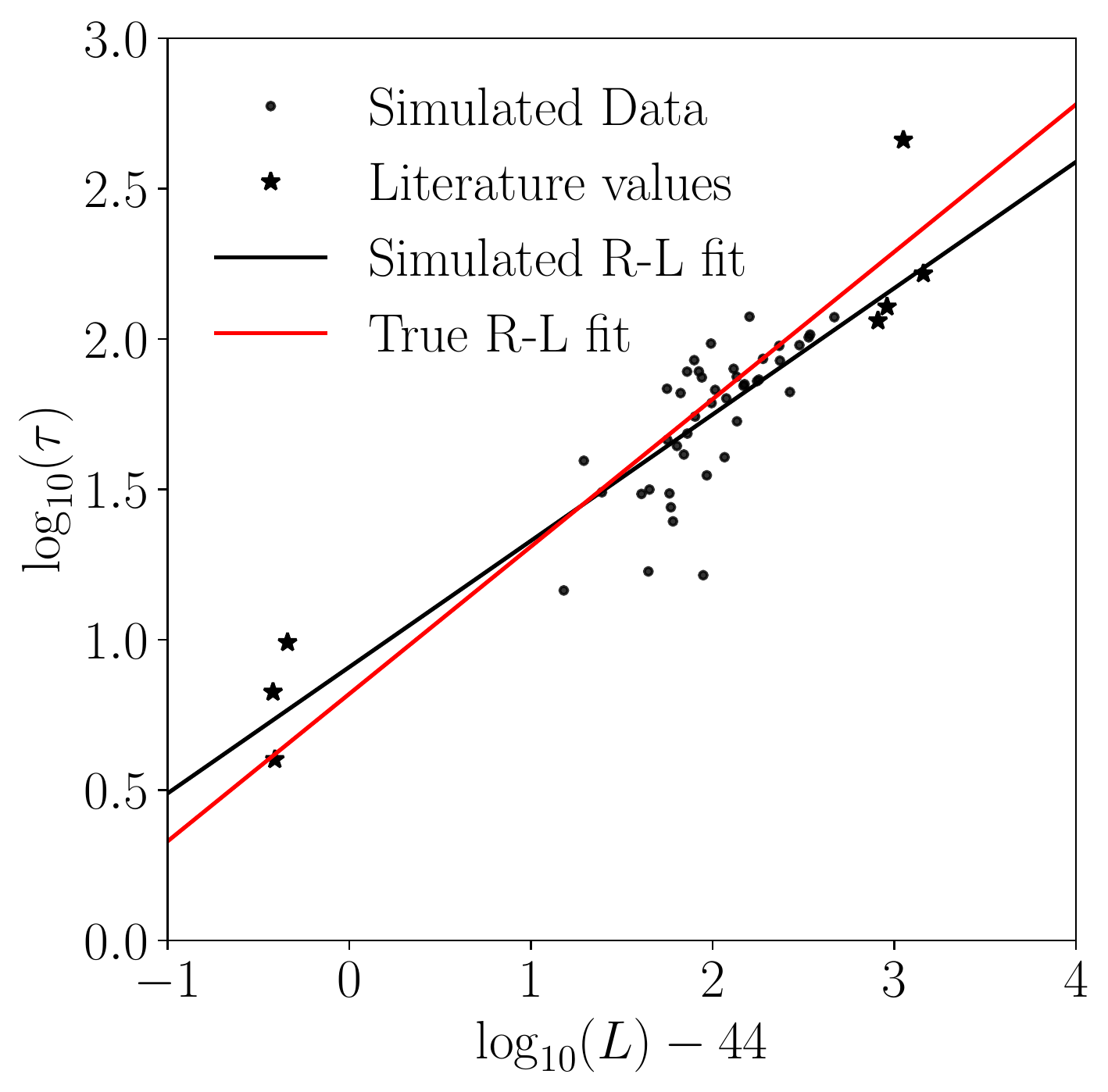}
    \caption{ An example of a best fit $\textit{R-L}$ relations computed using a subsample of sources. This figure shows the wide range of luminosities in the literature values that are used to supplement the smaller luminosity range of the simulated values.}
    \label{fig:RL_single}
\end{figure}

\begin{table}
    \centering
\caption{Rest frame time lags and 1350 \AA\ luminosities for all \ion{C}{iv} lags used to anchor the $\textit{R-L}$ relationship.}
	
	\begin{tabular}{|lcccccc}
		\hline
		AGN & $\log \lambda L_{\lambda}$ [ergs s$^{-1}$] &  $\tau_{\rm{RF}}$ [days] & Ref. \\
		\hline
        
		NGC 3783 & 43.59 $\pm$ 0.09  & $4.0^{+1.0}_{-1.5}$ &  1 \\[3pt]
        
		NGC 5548 Year 1 & 43.66 $\pm$ 0.14  & $9.8^{+1.9}_{-1.5}$ &  1 \\[3pt]
        
		NGC 5548 Year 5 & 43.58 $\pm$ 0.06  & $6.7^{+0.9}_{-1.0}$ &  1 \\[3pt]
        
		CT286 & 47.05 $\pm$ 0.12  & $459^{+71}_{-92}$ & 2 \\[3pt]
        
		CT406 & 46.91 $\pm$ 0.05  & $115^{+64}_{-86}$ &  2 \\[3pt]
        
		J214355 & 46.96 $\pm$ 0.07  & $128^{+91}_{-82}$ &  2\\[3pt]
        
		J221516 & 47.16 $\pm$ 0.12  & $165^{+98}_{-13}$ &  2\\
		\hline
	\end{tabular}
	\label{tab:existingLags}
	\\
	\flushleft
    \textbf{References:} (1) \citet{Peterson2005} and references therein; (2) \citet{Lira2018}.
\end{table}



\bsp	
\label{lastpage}
\end{document}